\documentclass[a4paper,11pt]{article}
\usepackage{jheppub} % for details on the use of the package, please see the JINST-author-manual
\usepackage{tikz}
\usepackage{tikz-cd}
\usetikzlibrary{decorations.markings}
\usetikzlibrary{shapes.geometric, arrows}
\tikzstyle{startstop} = [rectangle, rounded corners, minimum width=3cm, minimum height=0.6cm,text centered, draw=black]
\tikzstyle{process} = [rectangle, rounded corners, minimum width=3cm, minimum height=1cm, text centered, draw=black]
\tikzstyle{check} = [diamond, minimum width=3cm, minimum height=1cm, aspect=3.6, text centered, draw=black]
\tikzstyle{arrow} = [thick,->,>=stealth]

\usepackage{float}
\usepackage{longtable}
\usepackage{amsthm}
\usepackage{amsmath}
\allowdisplaybreaks
\newtheorem{theorem}{Theorem}[section]
\newtheorem{lemma}{Lemma}[section]

\newtheorem{corollary}{Corollary}[section]
\usepackage{lineno}
\usepackage{comment}
\title{\boldmath Application of the Meijer theorem in calculation of three-loop massive vacuum Feynman integrals and beyond }

\author[a,b]{Jian Wang,}
\author[a]{Dongyu Yang}

\affiliation[a]{School of Physics, Shandong University, Jinan, Shandong 250100, China}
\affiliation[b]{Center for High Energy Physics, Peking University, Beijing 100871, China}

\emailAdd{j.wang@sdu.edu.cn}
\emailAdd{202000141067@mail.sdu.edu.cn}

\abstract{We present an analytical method to calculate the three-loop massive Feynman integral in arbitrary dimensions. 
The method is based on the Mellin-Barnes representation of the Feynman integral.
The Meijer theorem and its corollary are used  to perform the integration over the Gamma functions, exponential functions, and hypergeometric functions. We also discuss the application of the method in other multi-loop Feynman integrals. }

\begin{document}
\maketitle
\flushbottom

\section{Introduction}
\label{sec:intro}

Multi-loop Feynman integrals play a crucial role in the application of quantum field theory.
They are indispensable in calculating precise scattering cross-sections for collider processes that are important for testing the standard model.
The massive vacuum loop integrals are one kind of the simplest multi-loop Feynman integrals.
The results of these integrals are useful in understanding the structure of multi-loop integrals, e.g. the classes of constants and functions that would appear for Feynman loop integrals. 
They also find applications in the calculation of effective potentials of some theories \cite{Martin:2015eia}.
Recently, the method of differential equations \cite{Kotikov:1990kg,Kotikov:1991pm} has proven to be efficient in modern loop calculation 
and the vacuum integrals can be taken as the boundaries of the differential equations that more complex integrals satisfy.

The two-loop vacuum integrals with arbitrary masses have been obtained in full analytic form without expansion in the space-time parameter $\epsilon = (4-D)/2$ \cite{Davydychev:1992mt}.
The three-loop single-scale vacuum integrals have been calculated up to the finite part $\mathcal{O}(\epsilon^0)$ \cite{Broadhurst:1991fi,Avdeev:1994db,Broadhurst:1998rz,Fleischer:1999mp,Chetyrkin:1999qi}, in which polylogarithms up to transcendental weight four are needed.
The results were extended to weight six in \cite{Lee:2010hs,Schroder:2005va,Kniehl:2017ikj}.
The analytical results of three-loop two-scale vacuum integrals were available up to  $\mathcal{O}(\epsilon^0)$ \cite{Bekavac:2009gz,Grigo:2012ji}.
A special three-loop two-scale vacuum integral was calculated with full dependence on $\epsilon$ \cite{Davydychev:2003mv}.
The finite parts of three-loop vacuum integrals with arbitrary mass pattern were computed numerically by using one- or two-dimensional integrals of elementary functions \cite{Freitas:2016zmy} or by solving the coupled first-order differential equations \cite{Martin:2016bgz}.
Numerical results for even higher loop vacuum bubble diagrams of a specific type have also been presented \cite{Groote:2005ay}.
Recently, the analytical results have been investigated with the Gel’fand-Kapranov-Zelevinsky hypergeometric systems \cite{Gu:2018aya,Gu:2020ypr,Zhang:2023fil,Zhang:2024mxd}.

In this work, we present an analytical calculation of a three-loop massive Feynman integral in arbitrary dimensions.
We adopted the Mellin-Barnes representation for the integral of Feynman parameters,
and applied the Meijer theorem and its corollary to perform the integration over the Gamma functions, exponential functions, and hypergeometric functions.
Our method is different from existing ones and we get a compact result of the three-loop vacuum integral.
We also discuss the application of our method in other Feynman integrals.

\section{Analytical calculation of the three-loop vacuum banana diagram}
\label{sec2}

In this section, we use the well-known three-loop vacuum banana diagram with four massive propagators, labeled by \textbf{BN}, to demonstrate our approach. 
In subsection \ref{sec2.1} we briefly review the definition of the Feynman integral of a three-loop vacuum diagram in $D$ dimensions and illustrate its Feynman parameter representation. 
In the following subsection \ref{sec2.2} we introduce the Mellin-Barnes transformation for the Symanzik polynomials so that the integration over the Feynman parameters is easily performed.
Then in subsection \ref{sec2.3}, we seek to lessen the number of parameters in Mellin-Barnes integrals 
and find it trivial to use Barnes's first lemma to minimize a six-fold complex integral into a four-dimensional one. 
After that, we have to deal with integrals of a combination of different kinds of functions, including exponential functions, $\Gamma$ functions, and hypergeometric functions. 
To perform the integration, 
we propose an application of the Meijer theorem and its corollary. 

\subsection{Feynman representation and graph polynomials}\label{sec2.1}
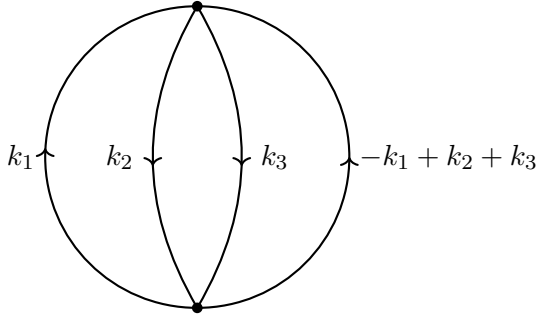
\begin{figure}[H]
    \centering
    \begin{tikzpicture}
\draw[-, bend left, thick,postaction={decorate, decoration={markings, mark=at position 0.5 with {\arrow{<}}}}]{} (0,-4) to (0,0) ;
\draw[-, bend right, thick,postaction={decorate, decoration={markings, mark=at position 0.5 with {\arrow{<}}}}]{} (0,-4) to (0,0) ;
\draw[-,thick, postaction={decorate, decoration={markings, mark=at position 0.5 with {\arrow{<}}}}]{} (0,0) arc (90:270:2) ;
\draw[-,thick,postaction={decorate, decoration={markings, mark=at position 0.5 with {\arrow{>}}}}]{} (0,-4) arc (-90:90:2) ;
\fill (0,0) circle (2pt);
\fill (0,-4) circle (2pt);
  
\node[left] at (-2,-2) {$k_1$};
\node[left] at (-0.7,-2) {$k_2$};
\node[right] at (0.7,-2) {$k_3$};
\node[right] at (2,-2) {$-k_1+k_2+k_3$};
\end{tikzpicture}
    \caption{The three-loop massive vacuum diagram \textbf{BN}.}
    \label{fig1}
\end{figure}
We begin our computation by writing down the normalized Feynman integral of the graph \textbf{BN} (see figure 
~\ref{fig1}):
\begin{equation}
    I\equiv e^{3\epsilon\gamma}\int[\mathrm{d}k_1][\mathrm{d}k_2][\mathrm{d}k_3]\frac{1}{D_1^{\nu_1}D_2^{\nu_2
    }D_3^{\nu_3}D_4^{\nu_4
    }}
\end{equation}
with the Euler-Mascheroni constant $\gamma=0.5772\cdots$, $\nu\equiv\nu_1+\nu_2+\nu_3+\nu_4$ and 
\begin{align}
 [\mathrm{d}k_i] & \equiv\frac{\mathrm{d}^Dk_i}{i\pi^{D/2}},\qquad i=1,2,3, \nonumber\\
     D_i & \equiv -k_i^2+m_i^2,\qquad i=1,2,3, \nonumber\\
     D_4 &\equiv -(-k_1+k_2+k_3)^2+m_4^2\,.
\end{align}
The corresponding Symanzik polynomials $\mathcal{U}$ and  $\mathcal{F}$ 
are given by \cite{BOGNER2_2010}
\begin{equation}\label{2.13}
    \mathcal{U}=a_1a_2a_3+a_1a_3a_4+a_2a_3a_4+a_1a_2a_4
\end{equation}
and
\begin{equation}\label{2.14}
    \mathcal{F}={(a_1a_2a_3+a_1a_3a_4+a_2a_3a_4+a_1a_2a_4)(a_1m_1^2+a_2m_2^2+a_3m_3^2+a_4m_4^2)}.
\end{equation}
The above integral is transformed to the integration over the Feynman parameters,
\begin{multline}
    I=\frac{e^{3\epsilon\gamma}\Gamma(\nu-\frac{3D}{2})}{\Gamma(\nu_1)\Gamma(\nu_2)\Gamma(\nu_3)\Gamma(\nu_4)}\int_0^1\mathrm{d}a_1\mathrm{d}a_2\mathrm{d}a_3\mathrm{d}a_4~\delta(1-a_1-a_2-a_3-a_4)\\
    \times a_1^{\nu_1-1}a_2^{\nu_2-1}a_3^{\nu_3-1}a_4^{\nu_4-1}\frac{(a_1a_2a_3+a_1a_3a_4+a_2a_3a_4+a_1a_2a_4)^{-\frac{D}{2}}}{(a_1m_1^2+a_2m_2^2+a_3m_3^2+a_4m_4^2)^{\nu-\frac{3D}{2}}}.
\end{multline}

\subsection{Mellin-Barnes representation}\label{sec2.2}
Making use of the Mellin-Barnes transformation~\cite{dubovyk2022mellinbarnes}
\begin{multline}
    (A_1+A_2+\cdots+A_n)^{-c}=\frac{1}{\Gamma(c)}\frac{1}{(2\pi i)^{n-1}}\int_{-i\infty}^{i\infty}\mathrm{d}\sigma_1\cdots\int_{-i\infty}^{i\infty}\mathrm{d}\sigma_{n-1}\\
    \Gamma(-\sigma_1)\cdots\Gamma(-\sigma_{n-1})\Gamma(\sigma_1+\cdots+\sigma_{n-1}+c)A_1^{\sigma_1}\cdots A_{n-1}^{\sigma_{n-1}}A_n^{-\sigma_1-\cdots\sigma_{n-1}-c},
\end{multline}
we derive 
\begin{multline}
    I=\frac{e^{3\epsilon\gamma}}{\Gamma(\nu_1)\Gamma(\nu_2)\Gamma(\nu_3)\Gamma(\nu_4)\Gamma(D/2)}\frac{1}{(2\pi i)^6}\times\\
    \int_{-i\infty}^{i\infty}\mathrm{d}\sigma_1\int_{-i\infty}^{i\infty}\mathrm{d}\sigma_2\int_{-i\infty}^{i\infty}\mathrm{d}\sigma_3 \int_{-i\infty}^{i\infty}\mathrm{d}\sigma_4\int_{-i\infty}^{i\infty}\mathrm{d}\sigma_5\int_{-i\infty}^{i\infty}\mathrm{d}\sigma_6\\
    \Gamma(-\sigma_1)\Gamma(-\sigma_2)\Gamma(-\sigma_3)\Gamma(\sigma_1+\sigma_2+\sigma_3+\frac{D}{2})\Gamma(-\sigma_4)\Gamma(-\sigma_5)\Gamma(-\sigma_6)\Gamma(\sigma_4+\sigma_5+\sigma_6+\nu-\frac{3D}{2})\times\\
    (m_1^2)^{\sigma_4}(m_2^2)^{\sigma_5}(m_3^2)^{\sigma_6}(m_4^2)^{-\sigma_4-\sigma_5-\sigma_6-\nu+\frac{3D}{2}}\int_0^1\mathrm{d}a_1\mathrm{d}a_2\mathrm{d}a_3\mathrm{d}a_4\cdot\delta(1-a_1-a_2-a_3-a_4)\times\\
    a_1^{\sigma_1+\sigma_2+\sigma_3+\sigma_4+\nu_1-1}a_2^{-\sigma_3+\sigma_5+\nu_2-\frac{D}{2}-1}a_3^{-\sigma_2+\sigma_6+\nu_3-\frac{D}{2}-1}a_4^{-\sigma_1-\sigma_4-\sigma_5-\sigma_6+\nu_4-\nu+D-1}.
\end{multline}
The integration over Feynman parameters can be done using the relation~\cite{dubovyk2022mellinbarnes}
\begin{equation}
    \int_0^1\bigg(\prod_{j=1}^{n}\mathrm{d}a_j\cdot a_j^{\nu_j-1}\bigg)\delta\bigg(1-\sum_{j=1}^{n}a_j\bigg)=\frac{\prod_{j=1}^n\Gamma(\nu_j)}{\Gamma(\nu_1+\cdots+\nu_n)}, 
\end{equation}
and we are left with
\begin{align}
    I  =& \frac{e^{3\epsilon\gamma}(m_4^2)^{-\nu+\frac{3D}{2}}}{\Gamma(\nu_1)\Gamma(\nu_2)\Gamma(\nu_3)\Gamma(\nu_4)\Gamma(D/2)}\frac{1}{(2\pi i)^6}\times \nonumber \\
    & \int_{-i\infty}^{i\infty}\mathrm{d}\sigma_1\int_{-i\infty}^{i\infty}\mathrm{d}\sigma_2\int_{-i\infty}^{i\infty}\mathrm{d}\sigma_3 \int_{-i\infty}^{i\infty}\mathrm{d}\sigma_4\int_{-i\infty}^{i\infty}\mathrm{d}\sigma_5\int_{-i\infty}^{i\infty}\mathrm{d}\sigma_6 \nonumber \\
    & \Gamma(-\sigma_1)\Gamma(-\sigma_2)\Gamma(-\sigma_3)\Gamma(\sigma_1+\sigma_2+\sigma_3+\frac{D}{2})\Gamma(-\sigma_4)\Gamma(-\sigma_5)\Gamma(-\sigma_6)
    \times \nonumber \\
    & \Gamma(\sigma_4+\sigma_5+\sigma_6+\nu-\frac{3D}{2}) \Gamma(\sigma_1+\sigma_2+\sigma_3+\sigma_4+\nu_1)\Gamma(-\sigma_3+\sigma_5+\nu_2-\frac{D}{2})\times \nonumber \\
    & \Gamma(-\sigma_2+\sigma_6+\nu_3-\frac{D}{2})\Gamma(-\sigma_1-\sigma_4-\sigma_5-\sigma_6+\nu_4-\nu+D)\times \nonumber \\
    & \bigg(\frac{m_1^2}{m_4^2}\bigg)^{\sigma_4}\bigg(\frac{m_2^2}{m_4^2}\bigg)^{\sigma_5}\bigg(\frac{m_3^2}{m_4^2}\bigg)^{\sigma_6}.
    \label{eq:generalBN}
\end{align}
The integration contours must be chosen such that the poles of $\Gamma(\cdots -\sigma)$ ($\Gamma(\cdots +\sigma)$) are to the right (left). 
For simplicity, we will focus on the case 
$m_1=m_2=m_3=m_4=1$ in the following calculation,
and discuss the general cases in the next section.

\subsection{Reduction of integration parameters}\label{sec2.3}
In order to reduce the number of integration parameters, we apply Barnes's first lemma~\cite{barnes1908new}.
\begin{lemma}[Barnes's first lemma]
    \begin{equation}\label{2.23}
    \frac{1}{2\pi i}\int_{-i\infty}^{i\infty}\mathrm{d}\sigma\Gamma(a+\sigma)\Gamma(b+\sigma)\Gamma(c-\sigma)\Gamma(d-\sigma)=\frac{\Gamma(a+c)\Gamma(a+d)\Gamma(b+c)\Gamma(b+d)}{\Gamma(a+b+c+d)}.
\end{equation}
\end{lemma}
\noindent The integral is convergent when $\textrm{Re}(a+b+c+d)<1$. However, this restriction can be removed after analytic continuation.

After integration over $\sigma_1$, $\sigma_4$, $\sigma_5$ and $\sigma_6$ in Eq. (\ref{eq:generalBN}), we obtain
\begin{align}\label{2.27}
    I  =& \frac{e^{3\epsilon\gamma}}{\Gamma(\nu_1)\Gamma(\nu_2)\Gamma(\nu_3)\Gamma(\nu_4)\Gamma(D/2)}\frac{1}{(2\pi i)^2}\int_{-i\infty}^{i\infty}\mathrm{d}\sigma_2\mathrm{d}\sigma_3 \nonumber \\
    & \Gamma(-\sigma_2)\Gamma(-\sigma_3)\Gamma(\sigma_2+\sigma_3+\frac{D}{2})\Gamma(\sigma_2+\sigma_3+\nu_1)\Gamma(\sigma_2+\sigma_3+\nu_4)\times \nonumber \\
    & \frac{\Gamma(\sigma_2+\sigma_3-\frac{D}{2}+\nu_1+\nu_4)\Gamma(-\sigma_2-\frac{D}{2}+\nu_3)\Gamma(-\sigma_3-\frac{D}{2}+\nu_2)}{\Gamma(2\sigma_2+2\sigma_3+\nu_1+\nu_4)}.
\end{align}
Without loss of generality, we set $\nu_1=\nu_2=\nu_3=\nu_4=1$ below. 
Notice that the argument of the Gamma function in the denominator contains $2\sigma_2$ and $2\sigma_3$.
They can be normalized by using the Gauss multiplication formula \cite{abramowitz1968handbook}.
\begin{theorem}[Gauss multiplication formula]\label{theorem2.1}
    For a positive integer $n$,
    \begin{equation}
        (2\pi)^{\frac{n-1}{2}}n^{\frac{1}{2}-nz}\Gamma(nz)=\prod_{k=0}^{n-1}\Gamma\bigg(z+\frac{k}{n}\bigg).
    \end{equation}
\end{theorem}

Then we have 
\begin{multline}\label{2.31}
    I=\frac{\sqrt{\pi}e^{3\epsilon\gamma}}{2\Gamma(2-\epsilon)}\frac{1}{(2\pi i)^2}\int_{-i\infty}^{i\infty}\mathrm{d}\sigma_2\mathrm{d}\sigma_3
    \Gamma(-\sigma_2)\Gamma(-\sigma_3)\Gamma(\sigma_2+\sigma_3+2-\epsilon)\Gamma(\sigma_2+\sigma_3+1)\times\\
    \frac{\Gamma(\sigma_2+\sigma_3+\epsilon)\Gamma(-\sigma_2-1+\epsilon)\Gamma(-\sigma_3-1+\epsilon)}{\Gamma(\sigma_2+\sigma_3+\frac{3}{2})}
    \cdot\Big(\frac{1}{4}\Big)^{\sigma_2}\cdot\Big(\frac{1}{4}\Big)^{\sigma_3}.
\end{multline}
The integration over $\sigma_2$ or $\sigma_3$ can be considered as a general integral form of the hypergeometric function and can be carried out according to the Meijer theorem~\cite{meijer1946}.
\begin{theorem}[Meijer Theorem]\label{theorem2.2}
    If the complex integral over s has a mixed product form of $\Gamma$ functions and exponential functions as the following 
    \begin{equation}
        \mathcal{I}(z)=\frac{1}{2\pi i}\int_{-i\infty}^{i\infty}\Gamma\Bigg[\begin{matrix}
            (a)+s,&(b)-s\\
            (c)+s,&(d)-s
        \end{matrix}\Bigg]z^s\mathrm{d}s,
    \end{equation}
    which is the Meijer G-function,
    then the result is given by
    \begin{multline}
        \Sigma_A(z)=\sum_{\mu=1}^Az^{-a_{\mu}}\Gamma\Bigg[\begin{matrix}
            (a)_\mu'-a_{\mu},&(b)+a_{\mu}\\
            (c)-a_{\mu},&(d)+a_{\mu}
        \end{matrix}\Bigg]\times\\
        {}_{B+C}F_{A+D-1}\Bigg[\begin{matrix}
            (b)+a_{\mu},1+a_{\mu}-(c);\\
            1+a_{\mu}-(a)_\mu',(d)+a_{\mu};
        \end{matrix}(-1)^{A+C}z^{-1}\Bigg]
    \end{multline}
    or
    \begin{multline}
        \Sigma_B(z)=\sum_{\nu=1}^Bz^{b_{\nu}}\Gamma\Bigg[\begin{matrix}
            (a)+b_{\nu},&(b)_\nu'-b_{\nu}\\
            (c)+b_{\nu},&(d)-b_{\nu}
        \end{matrix}\Bigg]\times\\
        {}_{A+D}F_{B+C-1}\Bigg[\begin{matrix}
            (a)+b_{\nu},1+b_{\nu}-(d);\\
            1+b_{\nu}-(b)_\nu',(c)+b_{\nu};
        \end{matrix}(-1)^{B+D}z\Bigg],
    \end{multline}
    where
    \begin{equation}
        \Gamma\Bigg[\begin{matrix}
            (a)+b_{\nu},&(b)_\nu'-b_{\nu}\\
            (c)+b_{\nu},&(d)-b_{\nu}
        \end{matrix}\Bigg]\equiv\frac{\Gamma((a)+b_{\nu})\Gamma((b)_\nu'-b_{\nu})}{\Gamma((c)+b_{\nu})\Gamma((d)-b_{\nu})}
    \end{equation}
    and (a) is defined by a set of constants or parameters independent of s, i.e., 
    \begin{equation}
        (a)\equiv \{ a_1, a_2,\dots,a_A \}.
    \end{equation}
    Meanwhile, the notation $(a)_\nu'$ represents a subset in which $a_{\nu}$ is excluded, i.e.
    \begin{equation}
        (a)_\nu'\equiv \{ a_1,\dots, a_{\nu-1},a_{\nu+1},\dots, a_A\} ,
    \end{equation}
    and $A\in\mathbb{N} $ is the number of elements included in (a).
     
   The choice between $\Sigma_A(z)$ and $\Sigma_B(z)$ is determined by a branch selection criterion:
    \begin{enumerate}
        \item[(i)] $\mathcal{I}(z)=\Sigma_A(z)\quad$ when $\quad B+C<A+D$ \quad and \quad $\frac{1}{2}\pi|A+B-C-D|>|\arg z|$
        
        or $\quad B+C=A+D\quad$ and $\quad|z|>1$.
        \item[(ii)] $\mathcal{I}(z)=\Sigma_B(z)\quad$ when $\quad B+C>A+D$ \quad and \quad $\frac{1}{2}\pi|A+B-C-D|>|\arg z|$
        
        or $\quad B+C=A+D\quad$ and $\quad|z|<1$.
            \item[(iii)] $\mathcal{I}(1)=\Sigma_A(1)=\Sigma_B(1)\quad$ when $\quad A-C=B-D\geq 0$ and $\textrm{Re}(\Sigma c+\Sigma d -\Sigma a- \Sigma b)>0$. 
    \end{enumerate}
\end{theorem}
After applying the Meijer theorem, eq. \eqref{2.31} is converted to
\begin{multline}\label{2.37}
    I=\frac{\sqrt{\pi }e^{3\epsilon\gamma}\Gamma(-1+\epsilon)}{2\Gamma(2-\epsilon)}\frac{1}{2\pi i}\int_{-i\infty}^{i\infty}\mathrm{d}\sigma_2\cdot\Gamma\Bigg[\begin{matrix}
        \sigma_2+2-\epsilon, \sigma_2+1, \sigma_2+\epsilon, -\sigma_2, -\sigma_2-1+\epsilon\\
        \sigma_2+\dfrac{3}{2}
    \end{matrix}\Bigg]\times\\
    {}_3F_2\Bigg[\begin{matrix}
        \sigma_2+2-\epsilon, \sigma_2+1, \sigma_2+\epsilon;\\
    2-\epsilon,\sigma_2+\dfrac{3}{2};
    \end{matrix}\quad\frac{1}{4}\Bigg]\bigg(\frac{1}{4}\bigg)^{\sigma_2}+\\
    \frac{\sqrt{\pi }e^{3\epsilon\gamma}\Gamma(1-\epsilon)}{2\Gamma(2-\epsilon)}\frac{1}{2\pi i}\cdot\bigg(\frac{1}{4}\bigg)^{-1+\epsilon}\int_{-i\infty}^{i\infty}\mathrm{d}\sigma_2\cdot\Gamma\Bigg[\begin{matrix}
        \sigma_2+1, \sigma_2+\epsilon, \sigma_2+2\epsilon-1, -\sigma_2, -\sigma_2-1+\epsilon\\
        \sigma_2+\dfrac{1}{2}+\epsilon
    \end{matrix}\Bigg]\times\\
    {}_3F_2\Bigg[\begin{matrix}
        \sigma_2-1+2\epsilon, \sigma_2+1, \sigma_2+\epsilon;\\
    \epsilon,\sigma_2+\dfrac{1}{2}+\epsilon;
    \end{matrix}\quad\frac{1}{4}\Bigg]\bigg(\frac{1}{4}\bigg)^{\sigma_2}.
\end{multline}
The integrands of eq. \eqref{2.37} appear to be a compound of $\Gamma$ functions, hypergeometric functions and exponential functions.
From the Meijer theorem, we derive the following corollary which helps to perform the integration\footnote{A similar form of this corollary was presented by L. J. Slater in ~\cite{slater1966generalized}. However, the result contains some typos.}.
The details of the proof are provided in appendix \ref{app:coro}.

\begin{corollary}
\label{corollary2.1}

    If the complex integral over s has a mixed product form of $\Gamma$ functions, exponential functions, and hypergeometric functions as the following 
    \begin{multline}
        \mathcal{I}(z)=\frac{1}{2\pi i}\int_{-i\infty}^{i\infty}\Gamma\Bigg[\begin{matrix}
            (a)+s,&(b)-s,&(g)+s,(h)-s\\
            (c)+s,&(d)-s,&(j)+s,(k)-s
        \end{matrix}\Bigg]\times\\
        {}_{A+B+E}F_{C+D+F}\Bigg[\begin{matrix}
            (a)+s,&(b)-s,& (e);\\
            (c)+s,&(d)-s,&(f);
        \end{matrix} ~x\Bigg] z^s\mathrm{d}s,
    \end{multline}
    then the result is given by
    \begin{multline}
        \Sigma_A(z)=\sum_{\mu=1}^{A}\Gamma\Bigg[\begin{matrix}
            (a)_\mu'-a_{\mu},&(b)+a_{\mu},&(g)-a_{\mu},&(h)+a_{\mu}\\
            (c)-a_{\mu},&(d)+a_{\mu},&(j)-a_{\mu},&(k)+a_{\mu}
        \end{matrix}\Bigg]\times\\
        \sum_{m=0}^{\infty}\sum_{n=0}^{\infty}\frac{((b)+a_{\mu})_{2m+n}((h)+a_{\mu})_{m+n}((e))_{m}(1+a_{\mu}-(c))_{n}(1+a_{\mu}-(j))_{m+n}}{(1+a_{\mu}-(a)_\mu')_{n}(1+a_{\mu}-(g))_{m+n}((f))_{m}((d)+a_{\mu})_{2m+n}((k)+a_{\mu})_{m+n}}\\
        \cdot\frac{x^m z^{-a_{\mu}-m-n}(-1)^{n(A+G-C-J)+m(G-J)}}{m!n!}\\
        +\sum_{\mu=1}^G\Gamma\Bigg[\begin{matrix}
            (a)-g_{\mu},&(b)+g_{\mu},&(g)_\mu'-g_{\mu},&(h)+g_{\mu}\\
            (c)-g_{\mu},&(d)+g_{\mu},&(j)-g_{\mu},&(k)+g_{\mu}
        \end{matrix}\Bigg]\times\\
        \sum_{m=0}^{\infty}\sum_{n=0}^{\infty}\frac{((a)-g_{\mu})_{m-n}((b)+g_{\mu})_{m+n}(1+g_{\mu}-(j))_n((h)+g_{\mu})_n((e))_m}{((c)-g_{\mu})_{m-n}((d)+g_{\mu})_{m+n}(1+g_{\mu}-(g)_\mu')_n((k)+g_{\mu})_n((f))_m}\\
        \cdot\frac{x^mz^{-g_{\mu}-n}(-1)^{n(G-J)}}{m!n!},
    \end{multline}
    or
    \begin{multline}
        \Sigma_B(z)=\sum_{\nu=1}^{B}\Gamma\Bigg[\begin{matrix}
            (a)+b_{\nu},&(b)_\nu'-b_{\nu},&(g)+ b_{\nu},&(h)-b_{\nu}\\
            (c)+b_{\nu},&(d)-b_{\nu},&(j)+b_{\nu},&(k)-b_{\nu}
        \end{matrix}\Bigg]\times\\
        \sum_{m=0}^{\infty}\sum_{n=0}^{\infty}\frac{((a)+b_{\nu})_{2m+n}((g)+b_{\nu})_{m+n}((e))_{m}(1+b_{\nu}-(d))_{n}(1+b_{\nu}-(k))_{m+n}}{(1+b_{\nu}-(b)_\nu')_{n}(1+b_{\nu}-(h))_{m+n}((f))_{m}((c)+b_{\nu})_{2m+n}((j)+b_{\nu})_{m+n}}\\
        \cdot\frac{x^m z^{b_{\nu}+m+n}(-1)^{n(B+H-K-D)+m(H-K)}}{m!n!}\\
        +\sum_{\nu=1}^H\Gamma\Bigg[\begin{matrix}
            (a)+h_{\nu},&(b)-h_{\nu},&(g)+ h_{\nu},&(h)_\nu'-h_{\nu}\\
            (c)+h_{\nu},&(d)-h_{\nu},&(j)+h_{\nu},&(k)-h_{\nu}
        \end{matrix}\Bigg]\times\\
        \sum_{m=0}^{\infty}\sum_{n=0}^{\infty}\frac{((a)+h_{\nu})_{m+n}((b)-h_{\nu})_{m-n}(1+h_{\nu}-(k))_n((g)+h_{\nu})_n((e))_m}{((d)-h_{\nu})_{m-n}((c)+h_{\nu})_{m+n}(1+h_{\nu}-(h)_\nu')_n((j)+h_{\nu})_n((f))_m}\\
        \cdot\frac{x^mz^{h_{\nu}+n}(-1)^{n(H-K)}}{m!n!},
    \end{multline}
    where the Pochhammer symbol $(a)_n$~\cite{knuth1992notes,abramowitz1968handbook} is defined by 
    \begin{equation}
        (a)_n\equiv \frac{\Gamma(a+n)}{\Gamma(a)},
    \end{equation}
    and the hypergeometric function ${}_{A+B+E}F_{C+D+F}(x)$ is absolutely and uniformly convergent in $x$.
    
    Provided that
        $\frac{1}{2}\pi|A+G+B+H-C-D-J-K|>|\arg z|$,
    the choice between $\Sigma_A(z)$ and $\Sigma_B(z)$ is determined by a branch selection criterion: 
    \begin{enumerate}
        \item[(i)]
        $\mathcal{I}(z)=\Sigma_A(z)$ \qquad
        when $\qquad A+G+D+K>B+H+C+J\qquad$

        or $\qquad A+G+D+K=B+H+C+J$ \qquad  and \qquad $|z|>1$.

        \item[(ii)] 
        $\mathcal{I}(z)=\Sigma_B(z)$ \qquad
        when $\qquad A+G+D+K<B+H+C+J\qquad$

        or $\qquad A+G+D+K=B+H+C+J$ \qquad and \qquad $|z|<1$.
 \end{enumerate}
\noindent 
Also, provided that $z=1$, and $\textrm{Re}(\Sigma c + \Sigma d + \Sigma j + \Sigma k - + \Sigma a - \Sigma b - \Sigma g - \Sigma h) > 0$, 
  \begin{enumerate}
 \item[(iii)]
        $\mathcal{I}(1)=\Sigma_A(1)=\Sigma_B(1)$\qquad 
        when $\qquad A+G-C-J=B+H-D-K\geq 0$. 
    \end{enumerate}
\end{corollary}

Finally, we obtain the analytical result for the three-loop  massive vacuum integral \textbf{BN} in arbitrary dimensions: 
\begin{multline}\label{219}
    I=\frac{\sqrt{\pi}e^{3\gamma\epsilon}\Gamma^2(-1+\epsilon)\Gamma(\epsilon)}{2\Gamma(\frac{3}{2})}\sum_{m=0}^{\infty}\sum_{n=0}^{\infty}\frac{(2-\epsilon)_{m+n}(1)_{m+n}(\epsilon)_{m+n}}{(2-\epsilon)_{n}(\frac{3}{2})_{m+n}(2-\epsilon)_m}\cdot\frac{1}{m!n!}\cdot\bigg(\frac{1}{4}\bigg)^{m+n}+\\
    \frac{\sqrt{\pi}e^{3\gamma\epsilon}\Gamma(\epsilon)\Gamma(-1+2\epsilon)\Gamma(1-\epsilon)\Gamma(-1+\epsilon)}{\Gamma(2-\epsilon)\Gamma(\frac{1}{2}+\epsilon)}\cdot\bigg(\frac{1}{4}\bigg)^{-1+\epsilon}\times\\
    \sum_{m=0}^{\infty}\sum_{n=0}^{\infty}\frac{(1)_{m+n}(\epsilon)_{m+n}(-1+2\epsilon)_{m+n}}{(\epsilon)_n(\frac{1}{2}+\epsilon)_{m+n}(2-\epsilon)_m}\cdot\frac{1}{m!n!}\cdot\bigg(\frac{1}{4}\bigg)^{m+n}+\\
    \frac{\sqrt{\pi}e^{3\gamma\epsilon}\Gamma(\epsilon)\Gamma(-1+2\epsilon)\Gamma(-2+3\epsilon)\Gamma^2(1-\epsilon)}{2\Gamma(2-\epsilon)\Gamma(-\frac{1}{2}+2\epsilon)}\cdot\bigg(\frac{1}{4}\bigg)^{-2+2\epsilon}\times\\
    \sum_{m=0}^{\infty}\sum_{n=0}^{\infty}\frac{(\epsilon)_{m+n}(-1+2\epsilon)_{m+n}(-2+3\epsilon)_{m+n}}{(\epsilon)_n(-\frac{1}{2}+2\epsilon)_{m+n}(\epsilon)_m}\cdot\frac{1}{m!n!}\cdot\bigg(\frac{1}{4}\bigg)^{m+n}.
\end{multline}
This result agrees with that in ref. \cite{Gu:2018aya} where a different method has been adopted.
The series of expansion around $\epsilon=0$ is given by 
\begin{multline}
    I=2\epsilon^{-3}+\frac{22}{3}\epsilon^{-2}+\bigg(\frac{27}{2}+8\ln 2+\frac{\pi^2}{2}-\frac{\partial^2}{\partial x\partial y}{}_2F_1\Big[\begin{matrix}
        x,\ y;\\
        -1/2;
    \end{matrix}\quad 1\Big]\Big|_{x=-1/2,y=-2}\\-2\frac{\partial^2}{\partial x\partial y}{}_2F_1\Big[\begin{matrix}
        -1/2,\ x;\\
        y;
    \end{matrix}\quad 1\Big]\Big|_{x=-2,y=-1/2} \bigg)\epsilon^{-1}+\mathcal{O}(\epsilon^0),
\end{multline}
which completely coincides with the numerical calculations in \cite{kniehl2017three}.

\section{Application of the method  in other Feynman integrals}\label{sec3.1}

The application of our calculation method in the above example is successful.
In this section, we present several more examples and a general algorithm for the calculation of  Feynman integrals.

\paragraph{BN diagrams with arbitrary masses}
In the above calculation, we have made an assumption that all masses of the four propagators are identical.
This does not mean that our method is only applicable in such a simple case.
Now we illustrate how to perform calculations for the \textbf{BN} diagrams with different masses.

Let us start from eq. (\ref{eq:generalBN}). 
It is easy to integrate over $\sigma_1$ using Barnes's first lemma, 
and we get
\begin{multline}
    I=\frac{e^{3\epsilon\gamma}(m_4^2)^{-\nu+\frac{3D}{2}}}{\Gamma(\nu_1)\Gamma(\nu_2)\Gamma(\nu_3)\Gamma(\nu_4)\Gamma(D/2)}\frac{1}{(2\pi i)^5}\times\\
    \int_{-i\infty}^{i\infty}\mathrm{d}\sigma_2\int_{-i\infty}^{i\infty}\mathrm{d}\sigma_3 \int_{-i\infty}^{i\infty}\mathrm{d}\sigma_4\int_{-i\infty}^{i\infty}\mathrm{d}\sigma_5\int_{-i\infty}^{i\infty}\mathrm{d}\sigma_6\bigg(\frac{m_1^2}{m_4^2}\bigg)^{\sigma_4}\bigg(\frac{m_2^2}{m_4^2}\bigg)^{\sigma_5}\bigg(\frac{m_3^2}{m_4^2}\bigg)^{\sigma_6}\times\\
    \Gamma(-\sigma_2)\Gamma(-\sigma_3)\Gamma(-\sigma_4)\Gamma(-\sigma_5)\Gamma(-\sigma_6)\Gamma(\sigma_4+\sigma_5+\sigma_6+\nu-\frac{3D}{2})\times\\
    \Gamma(-\sigma_3+\sigma_5+\nu_2-\frac{D}{2})\Gamma(-\sigma_2+\sigma_6+\nu_3-\frac{D}{2})\Gamma(\sigma_2+\sigma_3-\sigma_4-\sigma_5-\sigma_6+\frac{3D}{2}+\nu_4-\nu)\times\\
    \frac{\Gamma(\sigma_2+\sigma_3+\frac{D}{2})\Gamma(\sigma_2+\sigma_3+\sigma_4+\nu_1)\Gamma(\sigma_2+\sigma_3-\sigma_5-\sigma_6+D+\nu_4+\nu_1-\nu)}{\Gamma(2\sigma_2+2\sigma_3-\sigma_5-\sigma_6+\frac{3D}{2}+\nu_1+\nu_4-\nu)}.
\end{multline}
Assuming 
\begin{equation}
    m_1\geq\max(m_2,\ m_3,\ m_4),
\end{equation}
the  Meijer theorem can be applied to the integration of $\sigma_4$, yielding
\begin{multline}
    I=\frac{e^{3\epsilon\gamma}(m_4^2)^{-\nu+\frac{3D}{2}}}{\Gamma(\nu_1)\Gamma(\nu_2)\Gamma(\nu_3)\Gamma(\nu_4)\Gamma(D/2)}\frac{1}{(2\pi i)^4}\int_{-i\infty}^{i\infty}\mathrm{d}\sigma_2\mathrm{d}\sigma_3\mathrm{d}\sigma_5\mathrm{d}\sigma_6\\
    \times\Gamma(-\sigma_2)\Gamma(-\sigma_3)\Gamma(-\sigma_5)\Gamma(-\sigma_6)\Gamma(-\sigma_3+\sigma_5+\nu_2-D/2)\Gamma(-\sigma_2+\sigma_6+\nu_3-D/2)\\
    \times\frac{\Gamma(\sigma_2+\sigma_3+D/2)\Gamma(\sigma_2+\sigma_3-\sigma_5-\sigma_6+D+\nu_4+\nu_1-\nu)}{\Gamma(2\sigma_2+2\sigma_3-\sigma_5-\sigma_6+\frac{3D}{2}+\nu_1+\nu_4-\nu)}\bigg(\frac{m_2^2}{m_4^2}\bigg)^{\sigma_5}\bigg(\frac{m_3^2}{m_4^2}\bigg)^{\sigma_6}\\
    \times\Bigg\{\bigg(\frac{m_1^2}{m_4^2}\bigg)^{-\sigma_2-\sigma_3-\nu_1}\Gamma(-\sigma_2-\sigma_3+\sigma_5+\sigma_6+\nu-\nu_1-3D/2)\Gamma(\sigma_2+\sigma_3+\nu_1)\\
    \times\Gamma(2\sigma_2+2\sigma_3-\sigma_5-\sigma_6+3D/2+\nu_1+\nu_4-\nu)\\
    \times {}_{2}F_1\Bigg[\begin{matrix}
        \sigma_2+\sigma_3+\nu_1, 2\sigma_2+2\sigma_3-\sigma_5-\sigma_6+\frac{3D}{2}+\nu_1+\nu_4-\nu;\\
        \sigma_2+\sigma_3-\sigma_5-\sigma_6-\nu+\nu_1+\frac{3D}{2}+1;
    \end{matrix}\quad\frac{m_4^2}{m_1^2}\Bigg]\\
    +\bigg(\frac{m_1^2}{m_4^2}\bigg)^{-\sigma_5-\sigma_6-\nu+\frac{3D}{2}}\Gamma(\sigma_2+\sigma_3-\sigma_5-\sigma_6-\nu+\nu_1+3D/2)\Gamma(\sigma_5+\sigma_6+\nu-3D/2)\\
    \times\Gamma(\sigma_2+\sigma_3+\nu_4)_2F_1\Bigg[\begin{matrix}
        \sigma_5+\sigma_6+\nu-\frac{3D}{2},\ \sigma_2+\sigma_3+\nu_4;\\ 
        -\sigma_2-\sigma_3+\sigma_5+\sigma_6+\nu-\nu_1-\frac{3D}{2}+1;
    \end{matrix}\quad \frac{m_4^2}{m_1^2}\Bigg]\Bigg\}.
\end{multline}

It is manifest that our next step is to integrate over $\sigma_5$ (or $\sigma_6$) using Corollary \ref{corollary2.1}.
The result would contain Pochhammer symbols which are not suitable for the application of the Meijer theorem again. 
Therefore it is important to rewrite $\Sigma_A$ and $\Sigma_B$ in Corollary \ref{corollary2.1} in terms of $\Gamma$ functions: 
\begin{multline}
    \Sigma_A(z)=\sum_{\mu=1}^A\sum_{m=0}^\infty\sum_{n=0}^\infty\Gamma\Bigg[\begin{matrix}
(a)_\mu'-a_\mu,&1+a_\mu-(a)_\mu',&(g)-a_\mu,&1+a_\mu-(g),&(f),\\
(c)-a_\mu,&1+a_\mu-(c),&(j)-a_\mu,&1-a_\mu-(j),&(e),
    \end{matrix}\\
    \begin{matrix}
       (b)+a_\mu+2m+n,&(h)+a_\mu+m+n,&1+a_\mu-(c)+n,&1+ a_\mu-(j)+m+n,&(e)+m\\
       (d)+a_\mu+2m+n,&(k)+a_\mu+m+n,&1+a_\mu-(a)_\mu'+n,&1+a_\mu-(g)+m+n,&(f)+m
    \end{matrix}\Bigg]\\
    \times\frac{x^mz^{-a_\mu-m-n}(-1)^{n(A+G-C-J)+m(G-J)}}{m!n!}\\
    +\sum_{\mu=1}^G\sum_{m=0}^\infty\sum_{n=0}^\infty\Gamma\Bigg[\begin{matrix}
        (g)_\mu'-g_\mu,&1+g_\mu-(g)_\mu',&(f),&(a)-g_\mu+m-n,\\
        (j)-g_\mu,&1+g_\mu-(j),&(e),&(c)-g_\mu+m-n,
    \end{matrix}\\
    \begin{matrix}
        (b)+g_\mu+m+n,&1+g_\mu-(j)+n,&(h)+g_\mu+n,&(e)+m\\
        (d)+g_\mu+m+n,&1+g_\mu-(g)_\mu'+n,&(k)+g_\mu+n,&(f)+m
    \end{matrix}\Bigg]\frac{x^mz^{-g_\mu-n}(-1)^{n(G-J)}}{m!n!},
\end{multline}
\begin{multline}
    \Sigma_B(z)=\sum_{\nu=1}^B\sum_{m=0}^\infty\sum_{n=0}^\infty\Gamma\Bigg[\begin{matrix}
        (b)_\nu'-b_\nu,&(h)-b_\nu,&1+b_\nu-(b)_\nu',&1+b_\nu-(h),&(f),\\
        (d)-b_\nu,&(k)-b_\nu,&1+b_\nu-(d),&1+b_\nu-(k),&(e),
    \end{matrix}\\
    \begin{matrix}
        (a)+b_\nu+2m+n,&(g)+b_\nu+m+n,&(e)+m,&1+b_\nu-(d)+n,&1+b_\nu-(k)+m+n\\
        (c)+b_\nu+2m+n,&(j)+b_\nu+m+n,&(f)+m,&1+b_\nu-(b)_\nu'+n,&1+b_\nu-(h)+m+n
    \end{matrix}\Bigg]\\
    \times\frac{x^mz^{b_\nu+m+n}(-1)^{n(B+H-K-D)+m(H-K)}}{m!n!}\\
    +\sum_{\nu=1}^H\sum_{m=0}^\infty\sum_{n=0}^\infty\Gamma\Bigg[\begin{matrix}
        (h)_\nu'-h_\nu,&1+h_\nu-(h)_\nu',&(f),&(a)+h_\nu+m+n,&(b)-h_\nu+m-n,\\
        (k)-h_\nu,&1+h_\nu-(k),&(e),&(c)+h_\nu+m+n,&(d)-h_\nu+m-n,
    \end{matrix}\\
    \begin{matrix}
        (g)+h_\nu+n,&1+h_\nu-(k)+n,&(e)+m\\
        (j)+h_\nu+n,&1+h_\nu-(h)_\nu'+n,&(f)+m
    \end{matrix}\Bigg]\frac{x^mz^{h_\nu+n}(-1)^{n(H-K)}}{m!n!}.
\end{multline}

Then after the integration over $\sigma_5$, the result returns back to a strand of $\Gamma$-function summations,
\begin{multline}
    I=\frac{e^{3\epsilon\gamma}(m_4^2)^{-\nu+\frac{3D}{2}}}{\Gamma(\nu_1)\Gamma(\nu_2)\Gamma(\nu_3)\Gamma(\nu_4)\Gamma(D/2)}\sum_{m=0}^\infty\sum_{n=0}^\infty\frac{1}{m!n!}\bigg(\frac{m_4^2}{m_1^2}\bigg)^m\frac{1}{(2\pi i)^3}\iiint_{-i\infty}^{i\infty}\mathrm{d}\sigma_2\mathrm{d}\sigma_3\mathrm{d}\sigma_6\\
    \times\Gamma(-\sigma_2)\Gamma(-\sigma_3)\Gamma(-\sigma_6)\Gamma(-\sigma_2+\sigma_6+\nu_3-D/2)\Gamma(\sigma_2+\sigma_3+D/2)\\
    \times\Bigg\{\bigg(\frac{m_4^2}{m_1^2}\bigg)^{\sigma_{2,3}+\nu_1}\bigg(\frac{m_3^2}{m_4^2}\bigg)^{\sigma_3}\bigg(\frac{m_2^2}{m_4^2}\bigg)^{-n+\sigma_3-\nu_2+\frac{D}{2}}\Gamma\Bigg[\begin{matrix}
        -\sigma_2+\sigma_6-\nu_{1,2}+\nu-D,\\
        \sigma_2-\sigma_6+\nu_{1,2,4}-\nu+\frac{D}{2}+n,
    \end{matrix}\\
    \begin{matrix}
        2\sigma_2+\sigma_3-\sigma_6-\nu_3+D+m+n,&\sigma_{2,3}+\nu_1+m,&\sigma_2-\sigma_6+\nu_{1,2}-\nu+D+1\\
        2\sigma_2+\sigma_3-\sigma_6-\nu_3+D+n,&\sigma_3+\nu_2-\frac{D}{2}+n,&\sigma_2-\sigma_6+\nu_{1,2}-\nu+D+1+m+n
    \end{matrix}\Bigg]\\
    +\bigg(\frac{m_4^2}{m_1^2}\bigg)^{\sigma_{2,3}+\nu_1}\bigg(\frac{m_3^2}{m_4^2}\bigg)^{\sigma_6}\bigg(\frac{m_2^2}{m_4^2}\bigg)^{-n+\sigma_{2,3}-\sigma_6+\nu_1-\nu+\frac{3D}{2}}\Gamma\Bigg[\begin{matrix}
        n\\
        m+n,\ \nu_4-\frac{D}{2}+n
    \end{matrix}\Bigg]\\
    \times\Gamma\Bigg[\begin{matrix}
        \sigma_{2,3}+\nu_4+m+n,&\sigma_{2,3}+\nu_1+m,&-\sigma_2+\sigma_6-\nu_{1,2}+\nu_D+1,\\
        \sigma_{2,3}+\nu_4+n,&-\sigma_2+\sigma_6-\nu_{1,2}+\nu-D+1+n,&
    \end{matrix}\\
    \begin{matrix}
        \sigma_2-\sigma_6+\nu_{1,2}-\nu+D\\ -\sigma_{2,3}+\sigma_6-\nu_1+\nu-\frac{3D}{2}+n
    \end{matrix}\Bigg]\\
    +\bigg(\frac{m_4^2}{m_1^2}\bigg)^{\nu-\frac{3D}{2}}\bigg(\frac{m_3^2}{m_1^2}\bigg)^{\sigma_6}\bigg(\frac{m_2^2}{m_1^2}\bigg)^n\Gamma\Bigg[\begin{matrix}
        \sigma_{2,3}-\sigma_6+\nu_{1,4}-\nu+D,\\
        2\sigma_{2,3}-\sigma_6+\nu_{1,4}-\nu+\frac{3D}{2},
    \end{matrix}\\
    \begin{matrix}
    \\\sigma_{2,3}-\sigma_6+\nu_1-\nu+\frac{3D}{2},&-\sigma_{2,3}+\sigma_6-\nu_{1,4}+\nu-D+1,\\
        -2\sigma_{2,3}+\sigma_6-\nu_{1,4}+\nu-\frac{3D}{2}+1,&-\sigma_{2,3}+\sigma_6-\nu_1+\nu-\frac{3D}{2}+1+m+n,
    \end{matrix}\\
    \begin{matrix}
        -\sigma_{2,3}+\sigma_6-\nu_1+\nu-\frac{3D}{2}+1,&\sigma_6+\nu-\frac{3D}{2}+m+n,\\
        -\sigma_{2,3}+\sigma_6-\nu_{1,4}+\nu-D+1+n,&
    \end{matrix}\\
    \begin{matrix}
        -2\sigma_{2,3}+\sigma_6-\nu_{1,4}+\nu-\frac{3D}{2}+1+n,\ \sigma_{2,3}+\nu_4+m\\
        \ 
    \end{matrix}\Bigg]\\
    +\bigg(\frac{m_4^2}{m_1^2}\bigg)^{\nu-\frac{3D}{2}}\bigg(\frac{m_3^2}{m_1^2}\bigg)^{\sigma_6}\bigg(\frac{m_2^2}{m_1^2}\bigg)^{\sigma_{2,3}-\sigma_6+\nu_{1,4}-\nu+D+n}\Gamma\Bigg[\begin{matrix}
        -\nu_4+\frac{D}{2},\ \nu_4-\frac{D}{2}+1\\
        \nu_4-\frac{D}{2}+1+m+n
    \end{matrix}\Bigg]\\
    \times\Gamma\Bigg[\begin{matrix}
        -\sigma_{2,3}+\sigma_6-\nu_{1,4}+\nu-D,\ \sigma_{2,3}+\nu_4+m,\ \sigma_{2,3}-\sigma_6+\nu_{1,4}-\nu+D+1,\\
        \sigma_{2,3}+\frac{D}{2},
    \end{matrix}\\
\begin{matrix}
    \sigma_{2,3}+\nu_{1,4}-\frac{D}{2}+m+n,\ \sigma_2-\sigma_6+\nu_{1,2,4}-\nu+\frac{D}{2}+n,\ -\sigma_{2,3}-\frac{D}{2}+1+n\\
    -\sigma_{2,3}-\frac{D}{2}+1,\ \sigma_{2,3}-\sigma_6+\nu_{1,4}-\nu+D+1+n
\end{matrix}\Bigg]\\
+\bigg(\frac{m_4^2}{m_1^2}\bigg)^{\nu-\frac{3D}{2}}\bigg(\frac{m_3^2}{m_1^2}\bigg)^{\sigma_6}\bigg(\frac{m_2^2}{m_1^2}\bigg)^{\sigma_{2,3}-\sigma_6+\nu_1-\nu+\frac{3D}{2}+n}\Gamma\Bigg[\begin{matrix}
    \nu_4-\frac{D}{2},&-\nu_4+\frac{D}{2}+1,&1+n\\
    1+m+n,&-\nu_4+\frac{D}{2}+1+n&
    \end{matrix}\Bigg]\\
    \Gamma\Bigg[\begin{matrix}
        -\sigma_{2,3}+\sigma_6-\nu_1+\nu-\frac{3D}{2},&\sigma_{2,3}-\sigma_6+\nu_1-\nu+\frac{3D}{2}+1,&\sigma_{2,3}+\nu_1+m+n,\\
        -\sigma_{2,3}-\nu_4+1,&\sigma_{2,3}+\nu_4,&
    \end{matrix}\\
    \begin{matrix}
        -\sigma_{2,3}-\nu_4+1+n,\ \sigma_{2,3}+\nu_4+m\\
        \sigma_{2,3}-\sigma_6+\nu_1-\nu+\frac{3D}{2}+1+n
    \end{matrix}\Bigg]\Bigg\},
\end{multline}
where the Meijer theorem can be applied again.
In the above equation, we have used the abbreviation $\sigma_{i,j,\cdots}=\sigma_i+\sigma_j+\cdots $.

Note that in some addends (i.e. the first and the third ones) the coefficients of $\sigma_2$ and $\sigma_3$ are not equal to 1 or -1. In this case, we can use Theorem \ref{theorem2.1} and imitate what we have done in eqs.  \eqref{2.27}-\eqref{2.31}.
We have checked that the remaining three-fold integration can be done following such a strategy.

\begin{figure}[H]
    \centering
    \begin{tikzpicture}
\draw[-,thick,postaction={decorate, decoration={markings, mark=at position 0.5 with {\arrow{<}}}}]{} (0,-2)--(-1.732,-3);
\draw[-,thick,postaction={decorate, decoration={markings, mark=at position 0.5 with {\arrow{<}}}}]{} (0,-2)--(1.732,-3);
\draw[-,thick,postaction={decorate, decoration={markings, mark=at position 0.5 with {\arrow{<}}}}]{} (0,-2)--(0,0);
\draw[-,thick]{} (0,0) arc (90:270:2) ;
\draw[-,thick]{} (0,-4) arc (-90:90:2) ;
\fill (0,0) circle (2pt);
\fill (-1.732,-3) circle (2pt);
\fill (1.732,-3) circle (2pt);
\fill (0,-2) circle (2pt);
\node[left] at (-0.4,-2.12) {$k_1-k_2$};
\node[right] at (0.4,-2.12) {$k_2-k_3$};
\node[right] at (0,-1) {$k_3-k_1$};
\node[left] at (-1.7,-1) {$k_1$};
\node[right] at (1.7,-1) {$k_3$};
\node[below] at (0,-4) {$k_2$};
\end{tikzpicture}
    \caption{The three-loop massive vacuum diagram $D_6$.}
    \label{fig2}
\end{figure}
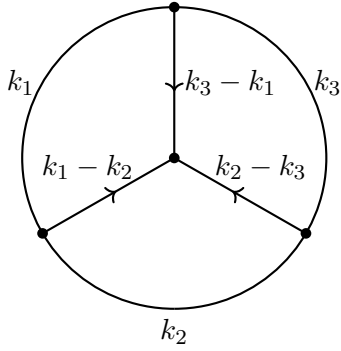

\paragraph{$D_6$ diagrams with six internal legs}
Another three-loop vacuum bubble worth discussion is the six-propagator $D_6$ diagram ~\cite{kniehl2017three,Broadhurst:1998rz,Freitas:2016zmy}, as shown in figure \ref{fig2}.
The definition of this integral is given by
\begin{equation}
    D_6\equiv e^{3\epsilon\gamma}\int\frac{[\mathrm{d}k_1][\mathrm{d}k_2][\mathrm{d}k_3]}{C_1^{\nu_1}C_2^{\nu_2}C_3^{\nu_3}C_4^{\nu_4}C_5^{\nu_5}C_6^{\nu_6}},
\end{equation}
where
\begin{equation}
    [\mathrm{d}k_i]\equiv\frac{\mathrm{d}^Dk_i}{i\pi^{D/2}},\qquad i=1,2,3
\end{equation}
and
\begin{align}
    C_i&\equiv -k_i^2+m_i^2,\qquad i=1,2,3,\\
    C_4&\equiv -(k_1-k_2)^2+m_4^2,\\
    C_5&\equiv -(k_2-k_3)^2+m_5^2,\\
    C_6&\equiv -(k_3-k_1)^2+m_6^2.
\end{align}
Its representations in Feynman parameters is
\begin{multline}
    D_6=\frac{e^{3\epsilon\gamma}\Gamma(\nu-\frac{3D}{2})}{\Gamma(\nu_1)\Gamma(\nu_2)\Gamma(\nu_3)\Gamma(\nu_4)\Gamma(\nu_5)\Gamma(\nu_6)}\int_0^1\mathrm{d}a_1\mathrm{d}a_2\mathrm{d}a_3\mathrm{d}a_4\mathrm{d}a_5\mathrm{d}a_6\\\delta\bigg(1-\sum_{i=1}^6a_i\bigg)\prod_{i=1}^6a^{\nu_i-1}
    \frac{\mathcal{A}^{-\frac{D}{2}}}{\mathcal{B}^{\nu-\frac{3D}{2}}},
\end{multline}
where
\begin{align}
    \begin{split}
        \mathcal{A}&=a_1a_2a_3+a_1a_3a_4+a_2a_3a_4+a_1a_2a_5+a_1a_3a_5+a_1a_4a_5+a_2a_4a_5+a_1a_3a_6\\
    &+a_2a_3a_6+a_1a_4a_6+a_2a_4a_6+a_3a_4a_6+a_1a_5a_6+a_2a_5a_6+a_3a_5a_6\\
    \mathcal{B}&=a_1m_1^2+a_2m_2^2+a_3m_3^2+a_4m_4^2+a_5m_5^2+a_6m_6^2.
    \end{split}
\end{align}
After introducing the Mellin-Barnes transformation to both the numerator and denominator, we have
\begin{multline}
    D_6=\frac{e^{3\epsilon\gamma}\Gamma(\nu-\frac{3D}{2})}{\Gamma(\nu_1)\Gamma(\nu_2)\Gamma(\nu_3)\Gamma(\nu_4)\Gamma(\nu_5)\Gamma(\nu_6)}\int_0^1\mathrm{d}a_1\mathrm{d}a_2\mathrm{d}a_3\mathrm{d}a_4\mathrm{d}a_5\mathrm{d}a_6\delta\bigg(1-\sum_{i=1}^6a_i\bigg)\\
    \times\frac{1}{(2\pi i)^{20}}\int_{-i\infty}^{i\infty}\mathrm{d}\sigma_1\cdots\mathrm{d}\sigma_{20}\prod_{i=1}^{20}\Gamma(-\sigma_i)\Gamma\bigg(\sum_{i=1}^{15}\sigma_i+\frac{D}{2}\bigg)\Gamma\bigg(\sum_{i=16}^{20}\sigma_i+\nu-\frac{3D}{2}\bigg)\\
    \times\bigg(\frac{m_1^2}{m_6^2}\bigg)^{\sigma_{16}}\bigg(\frac{m_2^2}{m_6^2}\bigg)^{\sigma_{17}}\bigg(\frac{m_3^2}{m_6^2}\bigg)^{\sigma_{18}}\bigg(\frac{m_4^2}{m_6^2}\bigg)^{\sigma_{19}}\bigg(\frac{m_5^2}{m_6^2}\bigg)^{\sigma_{20}}(m_6^2)^{-\nu+\frac{3D}{2}}\\
    \times a_1^{\sigma_{1,2,4,5,6,9,11,14,16}+\nu_1-1}a_2^{\sigma_{1,3,4,7,9,10,12,15,17}+\nu_2-1}
 a_3^{-\sigma_{4,6,7,9,11,12,14,15}+\sigma_{18}-\frac{D}{2}+\nu_3-1}\\
 \times a_4^{\sigma_{2,3,6,7,8,11,12,13,19}+\nu_4-1}
a_5^{-\sigma_{1,2,3,9,10,11,12,13}+\sigma_{20}-\frac{D}{2}+\nu_5-1} a_6^{-\sigma_{1,2,3,4,5,6,7,8,16,17,18,19,20}-\nu+D+\nu_6-1}.    
\end{multline}
 Integrating over Feynman parameters and setting $m_i=1$, we obtain the Mellin-Barnes  integral,
\begin{multline}
    D_6=\frac{e^{3\epsilon\gamma}\Gamma(\nu-\frac{3D}{2})}{\Gamma(\nu_1)\Gamma(\nu_2)\Gamma(\nu_3)\Gamma(\nu_4)\Gamma(\nu_5)\Gamma(\nu_6)}\frac{1}{(2\pi i)^{20}}\int_{-i\infty}^{i\infty}\mathrm{d}\sigma_1\cdots\mathrm{d}\sigma_{20}\prod_{i=1}^{20}\Gamma(-\sigma_i)\\
    \times \Gamma\bigg(\sum_{i=1}^{15}\sigma_i+\frac{D}{2}\bigg)\Gamma\bigg(\sum_{i=16}^{20}\sigma_i+\nu-\frac{3D}{2}\bigg)\Gamma(\sigma_{1,2,4,5,6,9,11,14,16}+\nu_1)\\
    \times\Gamma(\sigma_{1,3,4,7,9,10,12,15,17}+\nu_2) \Gamma(-\sigma_{4,6,7,9,11,12,14,15}+\sigma_{18}-\frac{D}{2}+\nu_3)\\
    \times\Gamma(\sigma_{2,3,6,7,8,11,12,13,19}+\nu_4)\Gamma(-\sigma_{1,2,3,9,10,11,12,13}+\sigma_{20}-\frac{D}{2}+\nu_5)\\\times\Gamma(-\sigma_{1,2,3,4,5,6,7,8,16,17,18,19,20}-\nu+D+\nu_6).
\end{multline}
The form of the integrand becomes obviously the product of $\Gamma$ functions and exponential functions again, though there are 20 Mellin-Barnes parameters.
Hence this calculation can be done by applying the Meijer theorem and its corollary in turns. 
For instance, if we integrate over $\sigma_1$, we obtain
\begin{multline}
    D_6=\frac{e^{3\epsilon\gamma}\Gamma(\nu-\frac{3D}{2})}{\Gamma(\nu_1)\Gamma(\nu_2)\Gamma(\nu_3)\Gamma(\nu_4)\Gamma(\nu_5)\Gamma(\nu_6)}\frac{1}{(2\pi i)^{19}}\int_{-i\infty}^{i\infty}\mathrm{d}\sigma_2\cdots\mathrm{d}\sigma_{20}\prod_{i=2}^{20}\Gamma(-\sigma_i)\\
    \times \Gamma\bigg(\sum_{i=16}^{20}\sigma_i+\nu-\frac{3D}{2}\bigg)\Gamma(-\sigma_{4,6,7,9,11,12,14,15}+\sigma_{18}-\frac{D}{2}+\nu_3)\Gamma(\sigma_{2,3,6,7,8,11,12,13,19}+\nu_4)\\
    \times\Bigg\{\Gamma\bigg(\sum_{i=2}^{15}\sigma_i+\frac{D}{2}\bigg)\Gamma(\sigma_{2,4,5,6,9,11,14,16}+\nu_1)\Gamma(\sigma_{3,4,7,9,10,12,15,17}+\nu_2)\\\times
    \Gamma(-\sigma_{2,3,9,10,11,12,13}+\sigma_{20}-\frac{D}{2}+\nu_5)\Gamma(-\sigma_{2,3,4,5,6,7,8,16,17,18,19,20}-\nu+D+\nu_6)\\
    \times{}_3F_2\bigg[\begin{matrix}
        \mathcal{A}_1,&\mathcal{A}_2,&\mathcal{A}_3;\\
        \mathcal{A}_4,&\mathcal{A}_5&;
    \end{matrix}\quad -1\bigg]\\
+\Gamma(\sigma_{4,5,6,7,8,14,15,20}+\nu_5)\Gamma(-\sigma_{3,10,12,13}+\sigma_{4,5,6,14,16,20}-\frac{D}{2}+\nu_1+\nu_5)\\
    \times\Gamma(-\sigma_{2,11,13}+\sigma_{4,7,15,17,20}-\frac{D}{2}+\nu_2+\nu_5)\Gamma(\sigma_{2,3,9,10,11,12,13}-\sigma_{20}+\frac{D}{2}-\nu_5)\\
\times\Gamma(\sigma_{4,5,6,7,8,16,17,19}-\sigma_{9,10,11,12,13}+2\sigma_{20}+\nu-\frac{3D}{2}-\nu_6+\nu_5)\\
\times{}_3F_2\bigg[\begin{matrix}
        \mathcal{B}_1,&\mathcal{B}_2,&\mathcal{B}_3;\\
        \mathcal{B}_4,&\mathcal{B}_5&;
    \end{matrix}\quad -1\bigg]\\
+\Gamma(\sigma_{9,10,11,12,13,14,15}-\sigma_{16,17,18,19,20}-\nu+\frac{3D}{2}+\nu_6)\Gamma(-\sigma_{3,7,8,17,18,19,20}+\sigma_{9,11,14}-\nu+D+\nu_1+\nu_6)\\
\times\Gamma(-\sigma_{2,5,6,8,16,18,19,20}+\sigma_{9,10,12,15}-\nu+D+\nu_2+\nu_6)\Gamma(\sigma_{2,3,4,5,6,7,8,,16,17,18,19,20}+\nu-D-\nu_6)\\
\times\Gamma(\sigma_{4,5,6,7,8,16,17,18,19}-\sigma_{9,10,11,12,13}+2\sigma_{20}+\nu-\frac{3D}{2}+\nu_5-\nu_6)\\
\times{}_3F_2\bigg[\begin{matrix}
        \mathcal{C}_1,&\mathcal{C}_2,&\mathcal{C}_3;\\
        \mathcal{C}_4,&\mathcal{C}_5&;
    \end{matrix}\quad -1\bigg]\Bigg\},
    \label{eq:D6res}
\end{multline}
where the explicit expressions of $\mathcal{A}_i, \mathcal{B}_i$, and $\mathcal{C}_i$ are collected in appendix \ref{app:ABC}.

The following computation becomes too tedious to be shown explicitly here,
though we do not expect any difficulties in principle.

\begin{figure}[H]
    \centering
    \begin{tikzpicture}
\draw[-,thick,postaction={decorate, decoration={markings, mark=at position 0.5 with {\arrow{>}}}}]{} (2,0)--(4,0);
\draw[-,thick,postaction={decorate, decoration={markings, mark=at position 0.5 with {\arrow{>}}}}]{} (-4,0)--(-2,0);
\draw[-,thick,postaction={decorate, decoration={markings, mark=at position 0.5 with {\arrow{>}}}}]{} (-2,0)--(2,0);
\draw[-,thick,postaction={decorate, decoration={markings, mark=at position 0.5 with {\arrow{<}}}}]{} (2,0) arc (0:180:2) ;
\draw[-,thick,postaction={decorate, decoration={markings, mark=at position 0.5 with {\arrow{>}}}}]{} (-2,0) arc (180:360:2) ;
\fill (-2,0) circle (2pt);
\fill (2,0) circle (2pt);
\fill (-4,0) circle (2pt);
\fill (4,0) circle (2pt);
\node[above] at (-3,0) {$p$};
\node[above] at (0,2) {$k_1$};
\node[above] at (0,-2) {$k_2$};
\node[above] at (0,0) {$p-k_1-k_2$};
\node[above] at (3,0) {$p$};

\end{tikzpicture}
    \caption{The two-loop sunset diagram.}
    \label{fig3}
\end{figure}
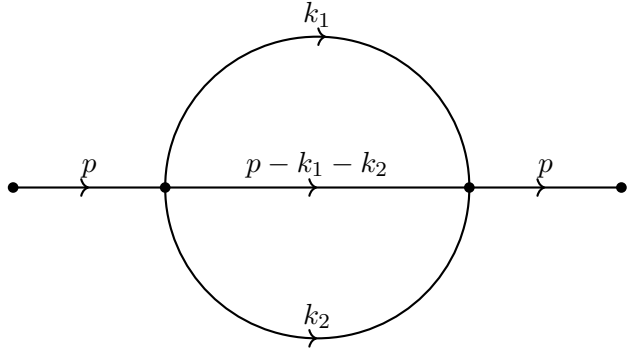

\paragraph{Diagrams with external legs}
Now, let us extend our exploration to the cases of diagrams with external legs. Take the well-known sunset diagram in figure \ref{fig3} as an example.
The Feynman integral is defined as
\begin{equation}
    S\equiv e^{2\epsilon\gamma}\iint\frac{[\mathrm{d}k_1][\mathrm{d}k_2]}{C_1^{\nu_1}C_2^{\nu_2}C_3^{\nu_3}},
\end{equation}
where 
\begin{align}
    [\mathrm{d}k_i]&\equiv\frac{\mathrm{d}^Dk_i}{i\pi^{D/2}},\qquad i=1,2,\\
    C_i&\equiv -k_i^2+m_i^2,\qquad i=1,2,\\
    C_3&\equiv -(-k_1-k_2+p)^2+m_3^2.
\end{align}
Adopting the Feynman parameterization, we have the following form,
\begin{multline}
    S=\frac{e^{2\epsilon\gamma}\Gamma(\nu-D)}{\Gamma(\nu_1)\Gamma(\nu_2)\Gamma(\nu_3)}\int_0^1\mathrm{d}a_1\mathrm{d}a_2\mathrm{d}a_3\delta(1-a_1-a_2-a_3)a_1^{\nu_1-1}a_2^{\nu_2-1}a_3^{\nu_3-1}\\
    \times\frac{(a_1a_2+a_1a_3+a_2a_3)^{\nu-\frac{3D}{2}}}{[(a_1a_2+a_1a_3+a_2a_3)(a_1m_1^2+a_2m_2^2+a_3m_3^2-a_3p^2)+a_3^2p^2(a_1+a_2)]^{\nu-D}}.
\end{multline}
Its corresponding Mellin-Barnes transformation is given by
\begin{multline}
    S=\frac{e^{2\epsilon\gamma}}{\Gamma(\nu_1)\Gamma(\nu_2)\Gamma(\nu_3)\Gamma(-\nu+\frac{3D}{2})}\frac{1}{(2\pi i)^8}\int_{-i\infty}^{i\infty}\mathrm{d}\sigma_1\mathrm{d}\sigma_2\mathrm{d}\sigma_3\mathrm{d}\sigma_4\mathrm{d}\sigma_5\mathrm{d}\sigma_6\mathrm{d}\sigma_7\mathrm{d}\sigma_8\\
    \times\Gamma(-\sigma_1)\Gamma(-\sigma_2)\Gamma(-\sigma_3)\Gamma(-\sigma_4)\Gamma(-\sigma_5)\Gamma(-\sigma_6)\Gamma(-\sigma_7)\Gamma(-\sigma_8)\Gamma(\sigma_{1,2}-\nu+\frac{3D}{2})\\
    \times\Gamma(\sigma_{3,4,5,6,7,8}+\nu-D)(m_1^2)^{\sigma_{3,5}}(m_2^2)^{\sigma_{4,7}}(m_3^2)^{\sigma_{6,8}}(m_1^2+m_2^2+m_3^2-p^2)^{-\sigma_{3,4,5,6,7,8}-\nu+D}\\
        \times\Gamma(\sigma_{1,2,3,5}-\sigma_{7,8}-\nu+D+\nu_1)\Gamma(-\sigma_{2,5,6}+\sigma_{4,7}+\nu_2-\frac{D}{2})\Gamma(-\sigma_{1,3,4}+\sigma_{6,8}+\nu_3-\frac{D}{2}).
\end{multline}
Again, we see that the integral is in the form ready for applying the Meijer theorem and its corollary. 
For example, let us perform the integration over $\sigma_1$ to demonstrate the feasibility of our algorithm, and the result reads
\begin{multline}
    S=\frac{e^{2\epsilon\gamma}}{\Gamma(\nu_1)\Gamma(\nu_2)\Gamma(\nu_3)\Gamma(-\nu+\frac{3D}{2})}\frac{1}{(2\pi i)^7}\int_{-i\infty}^{i\infty}\mathrm{d}\sigma_2\mathrm{d}\sigma_3\mathrm{d}\sigma_4\mathrm{d}\sigma_5\mathrm{d}\sigma_6\mathrm{d}\sigma_7\mathrm{d}\sigma_8\\
    \times\Gamma(-\sigma_2)\Gamma(-\sigma_3)\Gamma(-\sigma_4)\Gamma(-\sigma_5)\Gamma(-\sigma_6)\Gamma(-\sigma_7)\Gamma(-\sigma_8)\\
    \times\Gamma(\sigma_{3,4,5,6,7,8}+\nu-D)(m_1^2)^{\sigma_{3,5}}(m_2^2)^{\sigma_{4,7}}(m_3^2)^{\sigma_{6,8}}\\
    (m_1^2+m_2^2+m_3^2-p^2)^{-\sigma_{3,4,5,6,7,8}-\nu+D}\Gamma(-\sigma_{2,5,6}+\sigma_{4,7}+\nu_2-\frac{D}{2})\Gamma(\sigma_2-\nu+\frac{3D}{2})\\
   \times \Gamma(\sigma_{2,6,8}-\sigma_{3,4}-\nu+\nu_3+D)
    \frac{\Gamma(\sigma_{2,3,5}-\sigma_{7,8}-\nu+D+\nu_1)\Gamma(\sigma_{2,5,6}-\sigma_{4,7}-\nu_2+\frac{D}{2})}{\Gamma(2\sigma_2-\sigma_{4,7}+\sigma_{5,6}-\nu-\nu_2+2D)}.
\end{multline}
We do not bother to show the subsequent steps.

\begin{figure}[H]
    \centering
    \begin{tikzpicture}[node distance=2.6cm]

\node (start) [process] {Feynman Integrals};
\node (process1) [process, below of=start] {Mellin-Barnes Representation};
\node (check) [check, below of=process1] {Whether the calculation is done};
\node (process2) [process, below of=check] {\parbox{5cm}{Integration over $\Gamma$ functions and exponential functions}};
\node (process3) [process, below of=process2] {\parbox{5cm}{Integration over $\Gamma$ functions, exponential functions and hypergeometric functions}};
\node (process4) [process, below of=process3] {Analytical Results};
\node (process5) [process, below of=process4] {Asymptotic Series of $\epsilon$};
\node (compare) [process, below of=process5] {Numerical Results};

\draw[arrow](start) -- (process1){};
\draw[arrow](process1) -- (check){};
\draw [arrow] (check) -- node[anchor=west] {NO} (process2);
\draw [arrow] (process3) -- ++(7,0) |- node[anchor=south] {Corollary \ref{corollary2.1} \& Normalization} (check);
\draw [arrow] (check) -- ++(-5,0) |- node[anchor=north] {YES} (process4);
\draw [arrow] (process2) -- node[anchor=west] {Meijer Theorem  \& Normalization} (process3);
\draw [arrow] (process3) -- (process4);
\draw [arrow] (process4) -- node[anchor=west] {Asymptotic Expansion in $\epsilon$} (process5);
\draw [arrow] (process5) -- (compare);

\end{tikzpicture}
    \caption{The workflow of the calculation of Feynman integrals with the Meijer theorem and its corollary.  }
    \label{gross}
\end{figure}
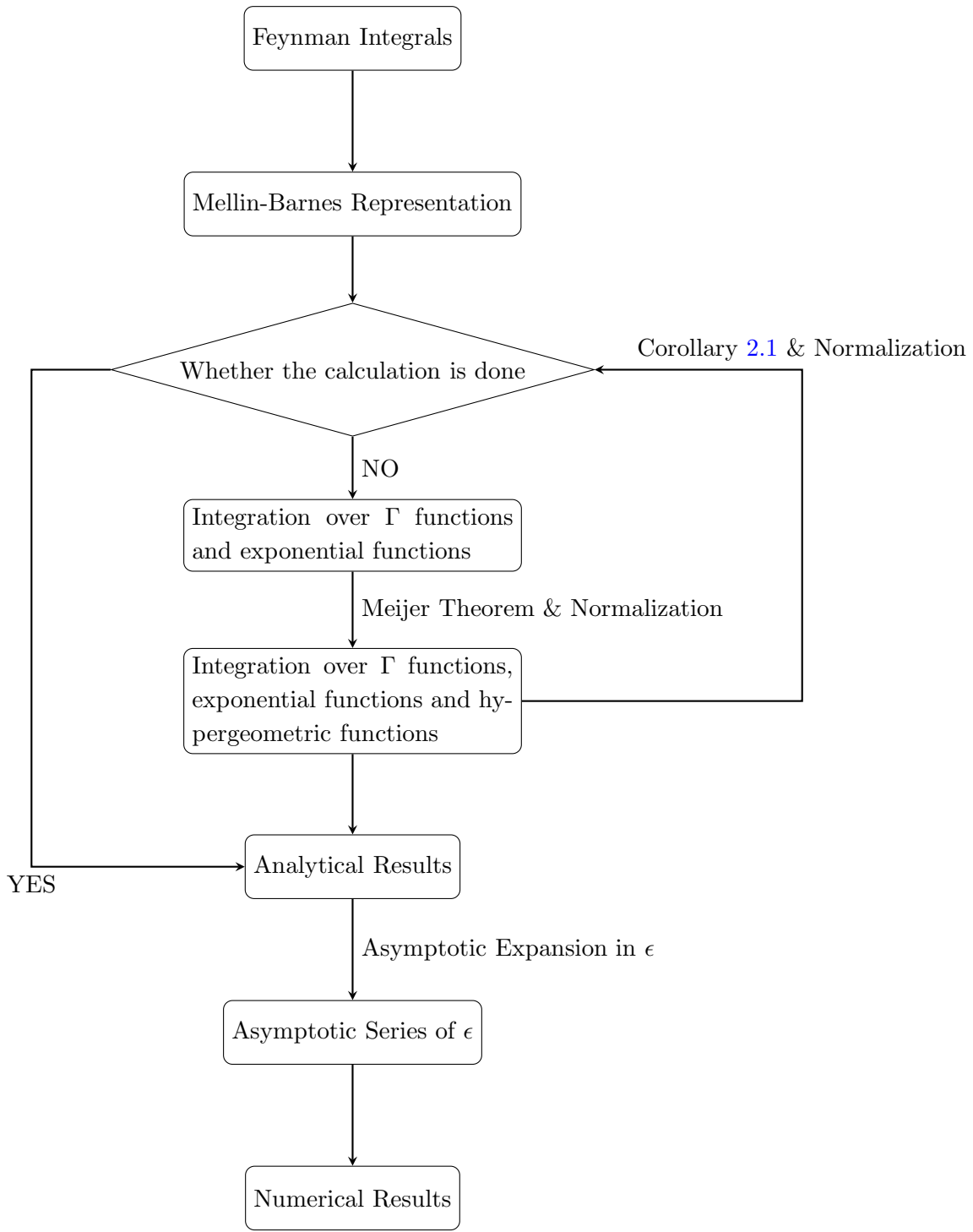

\paragraph{The calculation algorithm}
Summarizing the previous instances, we are ready to lay down a calculation algorithm for Feynman integral calculations. 
Starting with a given Feynman integral, we transform all the integral variables into Feynman parameters, and then to Mellin-Barnes forms. 
After these steps, the integrand would appear to be in the form of multiplication of $\Gamma$ functions and exponential functions, and it is natural to apply the Meijer theorem and express the result in products of $\Gamma$ functions, exponential functions, and hypergeometric functions.
The integration of these products can be dealt with by Corollary \ref{corollary2.1} and the result becomes integrals of $\Gamma$ functions and exponential functions again. 
The procedure will continue until the computation is eventually done. Note that before applying the theorems, it is essential to normalize the factors in front of the integration variables to $\pm 1$ by using Theorem \ref{theorem2.1}. 
A workflow is displayed in figure \ref{gross}.

\section{Conclusion and outlook}
We have proposed an analytical calculation method for the Feynman integrals in arbitrary dimensions by applying the Meijer theorem and its corollary, for which we have presented an explicit proof.
The feasibility has been demonstrated by the calculation of the three-loop vacuum banana diagram and discussed for several other multi-loop Feynman diagrams.
Even though our algorithm is general, there are still some obstacles arising in practical calculations. 
Here we list two of them.
\begin{itemize}
    \item \textbf{High computation complexity in multi-fold integration.}
It has been evident that in our discussion in section \ref{sec3.1} we completed only the first few steps to show the feasibility of our algorithm.
The expression starts to explode in the following steps.
An automatic computer program encoding the algorithm is needed.
\item \textbf{Necessity of normalization.}
The Meijer theorem and its corollary only apply to the integral in which the argument of the $\Gamma$ function is normalized.
In section \ref{sec2}, we have illustrated the normalization procedure using the theorem \ref{theorem2.1}.
However, one may encounter more complicated cases, such as $\Gamma(\sigma_3+\sigma_2)\Gamma(3\sigma_3+2\sigma_2)$ in the integrand.
The normalization of the $\sigma_2$ integration makes the $\sigma_3$ integration abnormal. 
Our method does not work in such cases.
\end{itemize}

\acknowledgments
 
This work was supported in part by the National Science Foundation of China (grant Nos. 12005117, 12321005, 12375076) and the Taishan Scholar Foundation of Shandong province (tsqn201909011).
The Feynman diagrams in this paper were drawn using the TikZ-Feynman package \cite{Ellis:2016jkw}.

\appendix

\section{The proof of Corollary \ref{corollary2.1}}\label{app:coro}

We begin with a careful consideration of the following complex integral
 \begin{multline}\label{b1}
        \mathcal{I}(z)=\frac{1}{2\pi i}\int_{-i\infty}^{i\infty}\Gamma\Bigg[\begin{matrix}
            (a)+s,&(b)-s,&(g)+s,(h)-s\\
            (c)+s,&(d)-s,&(j)+s,(k)-s
        \end{matrix}\Bigg]\times\\
        {}_{A+B+E}F_{C+D+F}\Bigg[\begin{matrix}
            (a)+s,&(b)-s,& (e);\\
            (c)+s,&(d)-s,&(f);
        \end{matrix}\quad x\Bigg]\quad z^s\mathrm{d}s.
    \end{multline}
    The generalized hypergeometric functions can be expanded as
    \begin{equation}
        {}_{A+B+E}F_{C+D+F}\Bigg[\begin{matrix}
            (a)+s,&(b)-s,& (e);\\
            (c)+s,&(d)-s,&(f);
        \end{matrix}\quad x\Bigg]\equiv\sum_{m=0}^\infty\frac{((a)+s)_m((b)-s)_m((e))_m}{((c)+s)_m((d)-s)_m((f))_m}\frac{x^m}{m!},
    \end{equation}
   or in terms of $\Gamma$ functions,
    \begin{multline}
        {}_{A+B+E}F_{C+D+F}\Bigg[\begin{matrix}
            (a)+s,&(b)-s,& (e);\\
            (c)+s,&(d)-s,&(f);
        \end{matrix}\quad x\Bigg]\\=\sum_{m=0}^\infty\Gamma\Bigg[\begin{matrix}
            (a)+s+m,&(b)-s+m,&(e)+m,&(c)+s,&(d)-s,&(f)\\
            (a)+s,&(b)-s,&(e),&(c)+s+m,&(d)-s+m,&(f)+m
        \end{matrix}\Bigg]\frac{x^m}{m!}.
    \end{multline}
    Then the previous integral~\eqref{b1} becomes
    \begin{multline}
        \mathcal{I}(z)=\sum_{m=0}^\infty\Gamma\Bigg[\begin{matrix}
            (e)+m,&(f)\\
            (e),&(f)+m
        \end{matrix}\Bigg]\frac{x^m}{m!}\frac{1}{2\pi i}\int_{-i\infty}^{i\infty}\mathrm{d}s z^{s}\\
        \times\Gamma\Bigg[\begin{matrix}
            (a)+s+m,&(b)-s+m,&(g)+s,&(h)-s\\
            (c)+s+m,&(d)-s+m,&(j)+s,&(k)-s
        \end{matrix}\Bigg]\,.
    \end{multline}
    According to the Meijer theorem, the result would be   
    \begin{enumerate}
        \item[(i)]
        $\mathcal{I}(z)=\Sigma_A(z)$
         when $\quad A+G+D+K>B+H+C+J\qquad$

        or $\quad A+G+D+K=B+H+C+J$ and $|z|>1$,
        \item[(ii)] 
        $\mathcal{I}(z)=\Sigma_B(z)$
        when $\quad A+G+D+K<B+H+C+J\qquad$

        or $\quad A+G+D+K=B+H+C+J$ and $|z|<1$,
        \end{enumerate}
        provided that     
        $\frac{1}{2}\pi|A+G+B+H-C-D-J-K|>|\arg z|$.
        In the special case of $z=1$ and $\textrm{Re}(\Sigma c + \Sigma d + \Sigma j + \Sigma k -  \Sigma a - \Sigma b - \Sigma g - \Sigma h) > 0$, 
    \begin{enumerate}
        \item[(iii)]
            $\mathcal{I}(1)=\Sigma_A(1)=\Sigma_B(1)$
        when $\quad A+G-C-J=B+H-D-K\geq 0$.
    \end{enumerate}

    The explicit expressions of $\Sigma_A$ and $\Sigma_B$ are given by
    \begin{multline}
        \Sigma_A(z)=\sum_{m=0}^\infty\Gamma\Bigg[\begin{matrix}
            (e)+m,&(f)\\
            (e),&(f)+m
        \end{matrix} \Bigg]\frac{x^m}{m!}\\
        \times\Bigg\{\sum_{\mu=1}^Az^{-a_\mu-m}\Gamma\Bigg[\begin{matrix}
        (a)_\mu'-a_{\mu},&(g)-a_\mu-m,&(b)+a_\mu+2m,&(h)+a_\mu+m\\
        (c)-a_\mu,&(j)-a_\mu-m,&(d)+a_\mu+2m,&(k)+a_\mu+m
        \end{matrix} \Bigg]\times\\
        {}_{B+H+C+J}F_{A+G+D+K-1}\Bigg[\begin{matrix}
            (b)+a_\mu+2m,&(h)+a_\mu+m,&1+a_\mu-(c),&1+a_\mu+m-(j);\\
            1+a_\mu-(a)_\mu',&1+a_\mu+m-(g),&(d)+a_\mu+2m,&(k)+a_\mu+m;
        \end{matrix}\quad \\(-1)^{A+G+C+J}z^{-1}\Bigg]\\
        +\sum_{\mu=1}^Gz^{-g_\mu}\Gamma\Bigg[\begin{matrix}
            (a)-g_\mu+m,&(g)_\mu'-g_\mu,&(b)+g_\mu+m&(h)+g_\mu\\
            (c)-g_\mu+m,&(j)-g_\mu,&(d)+g_\mu+m,&(k)+g_\mu
        \end{matrix} \Bigg]\times\\
        {}_{B+H+C+J}F_{A+G+D+K-1}\Bigg[ \begin{matrix}
            (b)+g_\mu+m,&(h)+g_\mu,&1+g_\mu-(c)-m,&1+g_\mu-(j);\\
            1+g_\mu-(a)-m,&1+g_\mu-(g)_\mu',&(d)+g_\mu+m,&(k)+g_\mu;
        \end{matrix}\\
        \quad (-1)^{A+G+C+J}z^{-1}\Bigg]\Bigg\}
    \end{multline}
    and
    \begin{multline}
        \Sigma_B(z)=\sum_{m=0}^\infty\Gamma\Bigg[\begin{matrix}
            (e)+m,&(f)\\
            (e),&(f)+m
        \end{matrix} \Bigg]\frac{x^m}{m!}\\
        \times\Bigg\{\sum_{\nu=1}^Bz^{b_\nu+m}\Gamma\Bigg[\begin{matrix}
        (a)+b_\nu+2m,&(g)+b_\nu+m,&(b)_\nu'-b_\nu,&(h)-b_\nu-m\\
        (c)+b_\nu+2m,&(j)+b_\nu+m,&(d)-b_\nu,&(k)-b_\nu-m
        \end{matrix} \Bigg]\times\\
        {}_{A+G+D+K}F_{B+H+C+J-1}\Bigg[\begin{matrix}
            (a)+b_\nu+2m,&(g)+b_\nu+m,&1+b_\nu-(d),&1+b_\nu+m-(k);\\
            1+b_\nu-(b)_\nu',&1+b_\nu+m-(h),&(c)+b_\nu+2m,&(j)+b_\nu+m;
        \end{matrix}\quad \\(-1)^{B+H+D+K}z\Bigg]\\
        +\sum_{\nu=1}^Hz^{h_\nu}\Gamma\Bigg[\begin{matrix}
            (a)+h_\nu+m,&(g)+h_\nu,&(b)-h_\nu+m,&(h)_\nu'-h_\nu\\
            (c)+h_\nu+m,&(j)+h_\nu,&(d)-h_\nu+m,&(k)-h_\nu
        \end{matrix} \Bigg]\times\\
        {}_{A+G+D+K}F_{B+H+C+J-1}\Bigg[ \begin{matrix}
            (a)+h_\nu+m,&(g)+h_\nu,&1-(d)+h_\nu-m,&1-(k)+h_\nu;\\
            1-(b)+h_\nu-m,&1-(h)_\nu'+h_\nu,&(c)+h_\nu+m,&(j)+h_\nu;
        \end{matrix}\\
        \quad (-1)^{B+H+D+K}z\Bigg]\Bigg\}\,.
    \end{multline}
    These results can be transformed into Phchhammer symbols with a more elegant form,
        \begin{multline}
        \Sigma_A(z)=\sum_{\mu=1}^{A}\Gamma\Bigg[\begin{matrix}
            (a)_\mu'-a_{\mu},&(b)+a_{\mu},&(g)-a_{\mu},&(h)+a_{\mu}\\
            (c)-a_{\mu},&(d)+a_{\mu},&(j)-a_{\mu},&(k)+a_{\mu}
        \end{matrix}\Bigg]\times\\
        \sum_{m=0}^{\infty}\sum_{n=0}^{\infty}\frac{((b)+a_{\mu})_{2m+n}((h)+a_{\mu})_{m+n}((e))_{m}(1+a_{\mu}-(c))_{n}(1+a_{\mu}-(j))_{m+n}}{(1+a_{\mu}-(a)_\mu')_{n}(1+a_{\mu}-(g))_{m+n}((f))_{m}((d)+a_{\mu})_{2m+n}((k)+a_{\mu})_{m+n}}\\
        \cdot\frac{x^m z^{-a_{\mu}-m-n}(-1)^{n(A+G-C-J)}}{m!n!}\\
        \times\frac{\Gamma((j)-a_\mu)\Gamma(1+a_\mu-(j))}{\Gamma(1+a_\mu-(j)+m)\Gamma((j)-a_\mu-m)}\cdot\frac{\Gamma(1+a_\mu-(g)+m)\Gamma((g)-a_\mu-m)}{\Gamma((g)-a_\mu)\Gamma(1+a_\mu-(g))}\\
        +\sum_{\mu=1}^G\Gamma\Bigg[\begin{matrix}
            (a)-g_{\mu},&(b)+g_{\mu},&(g)_\mu'-g_{\mu},&(h)+g_{\mu}\\
            (c)-g_{\mu},&(d)+g_{\mu},&(j)-g_{\mu},&(k)+g_{\mu}
        \end{matrix}\Bigg]\times\\
        \sum_{m=0}^{\infty}\sum_{n=0}^{\infty}\frac{((a)-g_{\mu})_{m-n}((b)+g_{\mu})_{m+n}(1+g_{\mu}-(j))_n((h)+g_{\mu})_n((e))_m}{((c)-g_{\mu})_{m-n}((d)+g_{\mu})_{m+n}(1+g_{\mu}-(g)_\mu')_n((k)+g_{\mu})_n((f))_m}\\
        \cdot\frac{x^mz^{-g_{\mu}-n}(-1)^{n(A+G-C-J)}}{m!n!}\\
        \times\frac{\Gamma((a)-g_\mu+m)\Gamma(1+g_\mu-(a)-m)}{\Gamma((a)-g_\mu+m-n)\Gamma(1+g_\mu-(a)-m+n)}\\
        \times\frac{\Gamma((c)-g_\mu+m-n)\Gamma(1+g_\mu-(c)-m+n)}{\Gamma((c)-g_\mu+m)\Gamma(1+g_\mu-(c)-m)},
    \end{multline}
    and
    \begin{multline}
        \Sigma_B(z)=\sum_{\nu=1}^{B}\Gamma\Bigg[\begin{matrix}
            (a)+b_{\nu},&(b)_\nu'-b_{\nu},&(g)+ b_{\nu},&(h)-b_{\nu}\\
            (c)+b_{\nu},&(d)-b_{\nu},&(j)+b_{\nu},&(k)-b_{\nu}
        \end{matrix}\Bigg]\times\\
        \sum_{m=0}^{\infty}\sum_{n=0}^{\infty}\frac{((a)+b_{\nu})_{2m+n}((g)+b_{\nu})_{m+n}((e))_{m}(1+b_{\nu}-(d))_{n}(1+b_{\nu}-(k))_{m+n}}{(1+b_{\nu}-(b)_\nu')_{n}(1+b_{\nu}-(h))_{m+n}((f))_{m}((c)+b_{\nu})_{2m+n}((j)+b_{\nu})_{m+n}}\\
        \cdot\frac{x^m z^{b_{\nu}+m+n}(-1)^{n(B+H-K-D)}}{m!n!}\\
        \times\frac{\Gamma((h)-b_\nu-m)\Gamma(1+b_\nu+m-(h))}{\Gamma(1+b_\nu-(h))\Gamma((h)-b_\nu)}\cdot\frac{\Gamma(1+b_\nu-(k))\Gamma((k)-b_\nu)}{\Gamma((k)-b_\nu-m)\Gamma(1+b_\nu+m-(k))}\\
        +\sum_{\nu=1}^H\Gamma\Bigg[\begin{matrix}
            (a)+h_{\nu},&(b)-h_{\nu},&(g)+ h_{\nu},&(h)_\nu'-h_{\nu}\\
            (c)+h_{\nu},&(d)-h_{\nu},&(j)+h_{\nu},&(k)-h_{\nu}
        \end{matrix}\Bigg]\times\\
        \sum_{m=0}^{\infty}\sum_{n=0}^{\infty}\frac{((a)+h_{\nu})_{m+n}((b)-h_{\nu})_{m-n}(1+h_{\nu}-(k))_n((g)+h_{\nu})_n((e))_m}{((d)-h_{\nu})_{m-n}((c)+h_{\nu})_{m+n}(1+h_{\nu}-(h)_\nu')_n((j)+h_{\nu})_n((f))_m}\\
        \cdot\frac{x^mz^{h_{\nu}+n}(-1)^{n(B+H-K-D)}}{m!n!}\\
        \times\frac{\Gamma((b)-h_\nu+m)\Gamma(1-(b)+h_\nu-m)}{\Gamma((b)-h_\nu+m-n)\Gamma(1-(b)+h_\nu-m+n)}\\
        \times\frac{\Gamma((d)-h_\nu+m-n)\Gamma(1-(d)+h_\nu-m+n)}{\Gamma((d)-h_\nu+m)\Gamma(1-(d)+h_\nu-m)}.
    \end{multline}
To simplify the redundant $\Gamma$ functions appearing in the last line of each summation series, we make use of Euler's reflection formula on $\Gamma$ functions.

    \begin{lemma}[Euler's reflection formula]
\begin{equation}
    \Gamma(1-z)\Gamma(z)=\frac{\pi}{\sin\pi z},\qquad z\notin\mathbb{Z}.
\end{equation}

    \end{lemma}
    A concise proof delivered by Dedekind can be found in ref.~\cite{srinivasan2011dedekind}. Replacing $z$ by $z-n$ in Euler's reflection formula and noting that
        \begin{equation}
            \frac{\pi}{\sin\pi(z-n)}=(-1)^n\frac{\pi}{\sin\pi z},
        \end{equation} 
    we derived another useful lemma.
    \begin{lemma}
        For $n\in\mathbb{Z}$ and $z\notin\mathbb{Z}$,
        \begin{equation}
            \frac{\Gamma(1-z)\Gamma(z)}{\Gamma(1+n-z)\Gamma(z-n)}=(-1)^n\,.
        \end{equation}
    \end{lemma}
    
    In our case, the redundant $\Gamma$ functions cancel out,
    \begin{align}
        \frac{\Gamma((j)-a_\mu)\Gamma(1+a_\mu-(j))}{\Gamma(1+a_\mu-(j)+m)\Gamma((j)-a_\mu-m)}&=(-1)^{mJ},\\
        \frac{\Gamma(1+a_\mu-(g)+m)\Gamma((g)-a_\mu-m)}{\Gamma((g)-a_\mu)\Gamma(1+a_\mu-(g))}&=(-1)^{mG},\\
        \frac{\Gamma((a)-g_\mu+m)\Gamma(1+g_\mu-(a)-m)}{\Gamma((a)-g_\mu+m-n)\Gamma(1+g_\mu-(a)-m+n)}&=(-1)^{nA},\\
        \frac{\Gamma((c)-g_\mu+m-n)\Gamma(1+g_\mu-(c)-m+n)}{\Gamma((c)-g_\mu+m)\Gamma(1+g_\mu-(c)-m)}&=(-1)^{nC},\\
        \frac{\Gamma((h)-b_\nu-m)\Gamma(1+b_\nu+m-(h))}{\Gamma(1+b_\nu-(h))\Gamma((h)-b_\nu)}&=(-1)^{mH},\\
        \frac{\Gamma(1+b_\nu-(k))\Gamma((k)-b_\nu)}{\Gamma((k)-b_\nu-m)\Gamma(1+b_\nu+m-(k))}&=(-1)^{mK},\\
        \frac{\Gamma((b)-h_\nu+m)\Gamma(1-(b)+h_\nu-m)}{\Gamma((b)-h_\nu+m-n)\Gamma(1-(b)+h_\nu-m+n)}&=(-1)^{nB},\\
        \frac{\Gamma((d)-h_\nu+m-n)\Gamma(1-(d)+h_\nu-m+n)}{\Gamma((d)-h_\nu+m)\Gamma(1-(d)+h_\nu-m)}&=(-1)^{nD}.
    \end{align}
    Therefore, we obtain 
        \begin{multline}
        \Sigma_A(z)=\sum_{\mu=1}^{A}\Gamma\Bigg[\begin{matrix}
            (a)_\mu'-a_{\mu},&(b)+a_{\mu},&(g)-a_{\mu},&(h)+a_{\mu}\\
            (c)-a_{\mu},&(d)+a_{\mu},&(j)-a_{\mu},&(k)+a_{\mu}
        \end{matrix}\Bigg]\times\\
        \sum_{m=0}^{\infty}\sum_{n=0}^{\infty}\frac{((b)+a_{\mu})_{2m+n}((h)+a_{\mu})_{m+n}((e))_{m}(1+a_{\mu}-(c))_{n}(1+a_{\mu}-(j))_{m+n}}{(1+a_{\mu}-(a)_\mu')_{n}(1+a_{\mu}-(g))_{m+n}((f))_{m}((d)+a_{\mu})_{2m+n}((k)+a_{\mu})_{m+n}}\\
        \cdot\frac{x^m z^{-a_{\mu}-m-n}(-1)^{n(A+G-C-J)+m(G-J)}}{m!n!}\\
        +\sum_{\mu=1}^G\Gamma\Bigg[\begin{matrix}
            (a)-g_{\mu},&(b)+g_{\mu},&(g)_\mu'-g_{\mu},&(h)+g_{\mu}\\
            (c)-g_{\mu},&(d)+g_{\mu},&(j)-g_{\mu},&(k)+g_{\mu}
        \end{matrix}\Bigg]\times\\
        \sum_{m=0}^{\infty}\sum_{n=0}^{\infty}\frac{((a)-g_{\mu})_{m-n}((b)+g_{\mu})_{m+n}(1+g_{\mu}-(j))_n((h)+g_{\mu})_n((e))_m}{((c)-g_{\mu})_{m-n}((d)+g_{\mu})_{m+n}(1+g_{\mu}-(g)_\mu')_n((k)+g_{\mu})_n((f))_m}\\
        \cdot\frac{x^mz^{-g_{\mu}-n}(-1)^{n(G-J)}}{m!n!},
    \end{multline}
    and
    \begin{multline}
        \Sigma_B(z)=\sum_{\nu=1}^{B}\Gamma\Bigg[\begin{matrix}
            (a)+b_{\nu},&(b)_\nu'-b_{\nu},&(g)+ b_{\nu},&(h)-b_{\nu}\\
            (c)+b_{\nu},&(d)-b_{\nu},&(j)+b_{\nu},&(k)-b_{\nu}
        \end{matrix}\Bigg]\times\\
        \sum_{m=0}^{\infty}\sum_{n=0}^{\infty}\frac{((a)+b_{\nu})_{2m+n}((g)+b_{\nu})_{m+n}((e))_{m}(1+b_{\nu}-(d))_{n}(1+b_{\nu}-(k))_{m+n}}{(1+b_{\nu}-(b)_\nu')_{n}(1+b_{\nu}-(h))_{m+n}((f))_{m}((c)+b_{\nu})_{2m+n}((j)+b_{\nu})_{m+n}}\\
        \cdot\frac{x^m z^{b_{\nu}+m+n}(-1)^{n(B+H-K-D)+m(H-K)}}{m!n!}\\
        +\sum_{\nu=1}^H\Gamma\Bigg[\begin{matrix}
            (a)+h_{\nu},&(b)-h_{\nu},&(g)+ h_{\nu},&(h)_\nu'-h_{\nu}\\
            (c)+h_{\nu},&(d)-h_{\nu},&(j)+h_{\nu},&(k)-h_{\nu}
        \end{matrix}\Bigg]\times\\
        \sum_{m=0}^{\infty}\sum_{n=0}^{\infty}\frac{((a)+h_{\nu})_{m+n}((b)-h_{\nu})_{m-n}(1+h_{\nu}-(k))_n((g)+h_{\nu})_n((e))_m}{((d)-h_{\nu})_{m-n}((c)+h_{\nu})_{m+n}(1+h_{\nu}-(h)_\nu')_n((j)+h_{\nu})_n((f))_m}\\
        \cdot\frac{x^mz^{h_{\nu}+n}(-1)^{n(H-K)}}{m!n!}.
    \end{multline}
which are the same results as in Corollary~\ref{corollary2.1}.

\section{The explicit expressions of $\mathcal{A}_i, \mathcal{B}_i$ and $\mathcal{C}_i$ in eq. (\ref{eq:D6res}). }
\label{app:ABC}

\begin{align}
    \mathcal{A}_1&=\sum_{i=2}^{15}\sigma_i+\frac{D}{2},\\
    \mathcal{A}_2&=\sigma_{2,4,5,6,9,11,14,16}+\nu_1,\\
    \mathcal{A}_3&=\sigma_{3,4,7,9,10,12,15,17}+\nu_2,\\
    \mathcal{A}_4&=1+\sigma_{2,3,9,10,11,12,13}-\sigma_{20}+\frac{D}{2}-\nu_5,\\
    \mathcal{A}_5&=1+\sigma_{2,3,4,5,6,7,8,16,17,18,19,20}+\nu-D-\nu_6,\\
    \mathcal{B}_1&=\sigma_{4,5,6,7,8,14,15,20}+\nu_5,\\
    \mathcal{B}_2&=-\sigma_{3,10,12,13}+\sigma_{4,5,6,14,16,20}-\frac{D}{2}+\nu_1+\nu_5,\\
    \mathcal{B}_3&=-\sigma_{2,11,13}+\sigma_{4,7,15,17,20}-\frac{D}{2}+\nu_2+\nu_5,\\
    \mathcal{B}_4&=1-\sigma_{2,3,9,10,11,12,13}+\sigma_{20}-\frac{D}{2}+\nu_5,\\
    \mathcal{B}_5&=1-\sigma_{4,5,6,7,8,16,17,19}+\sigma_{9,10,11,12,13}-2\sigma_{20}-\nu+\frac{3D}{2}+\nu_6-\nu_5,\\
\mathcal{C}_1&=\sigma_{9,10,11,12,13,14,15}-\sigma_{16,17,18,19,20}-\nu+\frac{3D}{2}+\nu_6,\\
    \mathcal{C}_2&=-\sigma_{3,7,8,17,18,19,20}+\sigma_{9,11,14}-\nu+D+\nu_1+\nu_6,\\
    \mathcal{C}_3&=-\sigma_{2,5,6,8,16,18,19,20}-+\sigma_{9,10,12,15}-\nu+D+\nu_2+\nu_6,\\
    \mathcal{C}_4&=1+\sigma_2-\sigma_{3,4,5,6,7,8,16,17,18,19,20}-\nu+D+\nu_6,\\
    \mathcal{C}_5&=1-\sigma_{4,5,6,7,8,16,17,18,19}+\sigma_{9,10,11,12,13}-2\sigma_{20}-\nu-\frac{3D}{2}-\nu_5+\nu_6.
\end{align}

% Bibliography

%% [A] Recommended: using JHEP.bst file
\bibliographystyle{JHEP}
\bibliography{biblio.bib}

\providecommand{\href}[2]{#2}\begingroup\raggedright\begin{thebibliography}{10}

\bibitem{Martin:2015eia}
S.P.~Martin, \emph{{Four-Loop Standard Model Effective Potential at Leading
  Order in QCD}}, \href{https://doi.org/10.1103/PhysRevD.92.054029}{\emph{Phys.
  Rev. D} {\bfseries 92} (2015) 054029}
  [\href{https://arxiv.org/abs/1508.00912}{{\ttfamily 1508.00912}}].

\bibitem{Kotikov:1990kg}
A.V.~Kotikov, \emph{{Differential equations method: New technique for massive
  Feynman diagrams calculation}},
  \href{https://doi.org/10.1016/0370-2693(91)90413-K}{\emph{Phys. Lett. B}
  {\bfseries 254} (1991) 158}.

\bibitem{Kotikov:1991pm}
A.V.~Kotikov, \emph{{Differential equation method: The Calculation of N point
  Feynman diagrams}},
  \href{https://doi.org/10.1016/0370-2693(91)90536-Y}{\emph{Phys. Lett. B}
  {\bfseries 267} (1991) 123}.

\bibitem{Davydychev:1992mt}
A.I.~Davydychev and J.B.~Tausk, \emph{{Two loop selfenergy diagrams with
  different masses and the momentum expansion}},
  \href{https://doi.org/10.1016/0550-3213(93)90338-P}{\emph{Nucl. Phys. B}
  {\bfseries 397} (1993) 123}.

\bibitem{Broadhurst:1991fi}
D.J.~Broadhurst, \emph{{Three loop on-shell charge renormalization without
  integration: Lambda-MS (QED) to four loops}},
  \href{https://doi.org/10.1007/BF01559486}{\emph{Z. Phys. C} {\bfseries 54}
  (1992) 599}.

\bibitem{Avdeev:1994db}
L.~Avdeev, J.~Fleischer, S.~Mikhailov and O.~Tarasov, \emph{{$0(\alpha
  \alpha_s^2)$ correction to the electroweak $\rho$ parameter}},
  \href{https://doi.org/10.1016/0370-2693(94)90573-8}{\emph{Phys. Lett. B}
  {\bfseries 336} (1994) 560}
  [\href{https://arxiv.org/abs/hep-ph/9406363}{{\ttfamily hep-ph/9406363}}].

\bibitem{Broadhurst:1998rz}
D.J.~Broadhurst, \emph{{Massive three - loop Feynman diagrams reducible to SC*
  primitives of algebras of the sixth root of unity}},
  \href{https://doi.org/10.1007/s100529900935}{\emph{Eur. Phys. J. C}
  {\bfseries 8} (1999) 311}
  [\href{https://arxiv.org/abs/hep-th/9803091}{{\ttfamily hep-th/9803091}}].

\bibitem{Fleischer:1999mp}
J.~Fleischer and M.Y.~Kalmykov, \emph{{Single mass scale diagrams: Construction
  of a basis for the epsilon expansion}},
  \href{https://doi.org/10.1016/S0370-2693(99)01321-0}{\emph{Phys. Lett. B}
  {\bfseries 470} (1999) 168}
  [\href{https://arxiv.org/abs/hep-ph/9910223}{{\ttfamily hep-ph/9910223}}].

\bibitem{Chetyrkin:1999qi}
K.G.~Chetyrkin and M.~Steinhauser, \emph{{The Relation between the MS-bar and
  the on-shell quark mass at order alpha(s)**3}},
  \href{https://doi.org/10.1016/S0550-3213(99)00784-1}{\emph{Nucl. Phys. B}
  {\bfseries 573} (2000) 617}
  [\href{https://arxiv.org/abs/hep-ph/9911434}{{\ttfamily hep-ph/9911434}}].

\bibitem{Lee:2010hs}
R.N.~Lee and I.S.~Terekhov, \emph{{Application of the DRA method to the
  calculation of the four-loop QED-type tadpoles}},
  \href{https://doi.org/10.1007/JHEP01(2011)068}{\emph{JHEP} {\bfseries 01}
  (2011) 068} [\href{https://arxiv.org/abs/1010.6117}{{\ttfamily 1010.6117}}].

\bibitem{Schroder:2005va}
Y.~Schroder and A.~Vuorinen, \emph{{High-precision epsilon expansions of
  single-mass-scale four-loop vacuum bubbles}},
  \href{https://doi.org/10.1088/1126-6708/2005/06/051}{\emph{JHEP} {\bfseries
  06} (2005) 051} [\href{https://arxiv.org/abs/hep-ph/0503209}{{\ttfamily
  hep-ph/0503209}}].

\bibitem{Kniehl:2017ikj}
B.A.~Kniehl, A.F.~Pikelner and O.L.~Veretin, \emph{{Three-loop massive tadpoles
  and polylogarithms through weight six}},
  \href{https://doi.org/10.1007/JHEP08(2017)024}{\emph{JHEP} {\bfseries 08}
  (2017) 024} [\href{https://arxiv.org/abs/1705.05136}{{\ttfamily
  1705.05136}}].

\bibitem{Bekavac:2009gz}
S.~Bekavac, A.G.~Grozin, D.~Seidel and V.A.~Smirnov, \emph{{Three-loop on-shell
  Feynman integrals with two masses}},
  \href{https://doi.org/10.1016/j.nuclphysb.2009.04.015}{\emph{Nucl. Phys. B}
  {\bfseries 819} (2009) 183}
  [\href{https://arxiv.org/abs/0903.4760}{{\ttfamily 0903.4760}}].

\bibitem{Grigo:2012ji}
J.~Grigo, J.~Hoff, P.~Marquard and M.~Steinhauser, \emph{{Moments of heavy
  quark correlators with two masses: exact mass dependence to three loops}},
  \href{https://doi.org/10.1016/j.nuclphysb.2012.07.007}{\emph{Nucl. Phys. B}
  {\bfseries 864} (2012) 580}
  [\href{https://arxiv.org/abs/1206.3418}{{\ttfamily 1206.3418}}].

\bibitem{Davydychev:2003mv}
A.I.~Davydychev and M.Y.~Kalmykov, \emph{{Massive Feynman diagrams and inverse
  binomial sums}},
  \href{https://doi.org/10.1016/j.nuclphysb.2004.08.020}{\emph{Nucl. Phys. B}
  {\bfseries 699} (2004) 3}
  [\href{https://arxiv.org/abs/hep-th/0303162}{{\ttfamily hep-th/0303162}}].

\bibitem{Freitas:2016zmy}
A.~Freitas, \emph{{Three-loop vacuum integrals with arbitrary masses}},
  \href{https://doi.org/10.1007/JHEP11(2016)145}{\emph{JHEP} {\bfseries 11}
  (2016) 145} [\href{https://arxiv.org/abs/1609.09159}{{\ttfamily
  1609.09159}}].

\bibitem{Martin:2016bgz}
S.P.~Martin and D.G.~Robertson, \emph{{Evaluation of the general 3-loop vacuum
  Feynman integral}},
  \href{https://doi.org/10.1103/PhysRevD.95.016008}{\emph{Phys. Rev. D}
  {\bfseries 95} (2017) 016008}
  [\href{https://arxiv.org/abs/1610.07720}{{\ttfamily 1610.07720}}].

\bibitem{Groote:2005ay}
S.~Groote, J.G.~Korner and A.A.~Pivovarov, \emph{{On the evaluation of a
  certain class of Feynman diagrams in x-space: Sunrise-type topologies at any
  loop order}}, \href{https://doi.org/10.1016/j.aop.2006.11.001}{\emph{Annals
  Phys.} {\bfseries 322} (2007) 2374}
  [\href{https://arxiv.org/abs/hep-ph/0506286}{{\ttfamily hep-ph/0506286}}].

\bibitem{Gu:2018aya}
Z.-H.~Gu and H.-B.~Zhang, \emph{{Three-loop vacuum integral with
  four-propagators using hypergeometry}},
  \href{https://doi.org/10.1088/1674-1137/43/8/083102}{\emph{Chin. Phys. C}
  {\bfseries 43} (2019) 083102}
  [\href{https://arxiv.org/abs/1811.10429}{{\ttfamily 1811.10429}}].

\bibitem{Gu:2020ypr}
Z.-H.~Gu, H.-B.~Zhang and T.-F.~Feng, \emph{{Hypergeometric expression for a
  three-loop vacuum integral}},
  \href{https://doi.org/10.1142/S0217751X2050089X}{\emph{Int. J. Mod. Phys. A}
  {\bfseries 35} (2020) 2050089}.

\bibitem{Zhang:2023fil}
H.-B.~Zhang and T.-F.~Feng, \emph{{GKZ hypergeometric systems of the three-loop
  vacuum Feynman integrals}},
  \href{https://doi.org/10.1007/JHEP05(2023)075}{\emph{JHEP} {\bfseries 05}
  (2023) 075} [\href{https://arxiv.org/abs/2303.02795}{{\ttfamily
  2303.02795}}].

\bibitem{Zhang:2024mxd}
H.-B.~Zhang and T.-F.~Feng, \emph{{GKZ hypergeometric systems of the four-loop
  vacuum Feynman integrals}},
  \href{https://arxiv.org/abs/2403.13025}{{\ttfamily 2403.13025}}.

\bibitem{BOGNER2_2010}
C.~Bogner and S.~Weinzierl, \emph{{Feynman graph polynomials}},
  \href{https://doi.org/10.1142/S0217751X10049438}{\emph{Int. J. Mod. Phys. A}
  {\bfseries 25} (2010) 2585}
  [\href{https://arxiv.org/abs/1002.3458}{{\ttfamily 1002.3458}}].

\bibitem{dubovyk2022mellinbarnes}
I.~Dubovyk, J.~Gluza and G.~Somogyi, \emph{{Mellin-Barnes Integrals: A Primer
  on Particle Physics Applications}},
  \href{https://doi.org/10.1007/978-3-031-14272-7}{\emph{Lect. Notes Phys.}
  {\bfseries 1008} (2022) pp.}
  [\href{https://arxiv.org/abs/2211.13733}{{\ttfamily 2211.13733}}].

\bibitem{barnes1908new}
E.W.~Barnes, \emph{A new development of the theory of the hypergeometric
  functions}, {\emph{Proceedings of the London Mathematical Society} {\bfseries
  2} (1908) 141}.

\bibitem{abramowitz1968handbook}
M.~Abramowitz and I.A.~Stegun, \emph{Handbook of mathematical functions with
  formulas, graphs, and mathematical tables}, vol.~55, US Government printing
  office (1968).

\bibitem{meijer1946}
C.~Meijer, \emph{{On the $G$-function}}, {\emph{Proc. Kon. Akad. v. Wetensch.}
  {\bfseries 49} (1946) 227}.

\bibitem{slater1966generalized}
L.J.~Slater, \emph{Generalized hypergeometric functions}, Cambridge University
  Press (1966).

\bibitem{knuth1992notes}
D.E.~Knuth, \emph{{Two Notes on Notation}},
  \href{https://doi.org/10.1080/00029890.1992.11995869}{\emph{The American
  Mathematical Monthly} {\bfseries 99} (1992) 403}
  [\href{https://arxiv.org/abs/math/9205211}{{\ttfamily math/9205211}}].

\bibitem{kniehl2017three}
B.A.~Kniehl, A.F.~Pikelner and O.L.~Veretin, \emph{{Three-loop massive tadpoles
  and polylogarithms through weight six}},
  \href{https://doi.org/10.1007/JHEP08(2017)024}{\emph{JHEP} {\bfseries 08}
  (2017) 024} [\href{https://arxiv.org/abs/1705.05136}{{\ttfamily
  1705.05136}}].

\bibitem{Ellis:2016jkw}
J.~Ellis, \emph{{TikZ-Feynman: Feynman diagrams with TikZ}},
  \href{https://doi.org/10.1016/j.cpc.2016.08.019}{\emph{Comput. Phys. Commun.}
  {\bfseries 210} (2017) 103}
  [\href{https://arxiv.org/abs/1601.05437}{{\ttfamily 1601.05437}}].

\bibitem{srinivasan2011dedekind}
G.K.~Srinivasan, \emph{{Dedekind’s proof of Euler’s reflection formula via
  ODEs}}, {\emph{Mathematics Newsletter} {\bfseries 21} (2011) 82}.

\end{thebibliography}\endgroup

%end{thebibliography}
\end{document}